\documentclass[longauth]{aa} % for the long lists of affiliations
\usepackage{graphicx}
\usepackage{txfonts}
\usepackage{hyperref}
\usepackage{colortbl}
\usepackage{xcolor}
\usepackage{multirow}
\begin{document}

%\title{GAPS XLVII: is V1298\,Tau the smoking gun of a Planet--Planet scattering event?
%\thanks{Based on observations made with ESO Telescopes at the La Silla Paranal Observatory under programmes ID 0104.C-0247(A) and 0108.C-0509(A) }}
\title{The GAPS program at TNG XLVII: The unusual formation history of V1298\,Tau.
\thanks{Based on observations made with the Italian Telescopio Nazionale Galileo (TNG) operated by the Fundaci\`on Galileo Galilei (FGG) of the Istituto Nazionale di Astrofisica (INAF) at the Observatorio del Roque de los Muchachos (La Palma, Canary Islands, Spain).}}
\author{D. Turrini\inst{\ref{ins:oato},\ref{ins:icsc},\ref{ins:iaps}}, F. Marzari\inst{\ref{ins:dfapd}}, D. Polychroni\inst{\ref{ins:oats}}, R. Claudi\inst{\ref{ins:oapd}}, S. Desidera\inst{\ref{ins:oapd}}, D. Mesa \inst{\ref{ins:oapd}}, M. Pinamonti\inst{\ref{ins:oato}}, A. Sozzetti\inst{\ref{ins:oato}}, A. Su\'{a}rez Mascare\~{n}o\inst{\ref{ins:iac},\ref{ins:unila}}, M. Damasso \inst{\ref{ins:oato}}, S. Benatti\inst{\ref{ins:oapa}}, L. Malavolta\inst{\ref{ins:dfapd}}, G. Micela \inst{\ref{ins:oapa}}, A. Zinzi \inst{\ref{ins:ssdc}}, 
V. J. S. B\'ejar \inst{\ref{ins:iac},\ref{ins:unila}},
K. Biazzo\inst{\ref{ins:oarm}}, 
A. Bignamini\inst{\ref{ins:oats}},
M. Bonavita \inst{\ref{ins:open},\ref{ins:supa},\ref{ins:oapd}}, 
F. Borsa \inst{\ref{ins:oabr}},
C. del Burgo \inst{\ref{ins:unimx}}, %D\'\i az 
G. Chauvin \inst{\ref{ins:cote}}, 
P. Delorme\inst{\ref{ins:ipag}},
J. I. Gonz\'alez Hern\'andez\inst{\ref{ins:iac},\ref{ins:unila}},
R. Gratton\inst{\ref{ins:oapd}},
J. Hagelberg\inst{\ref{ins:unige}},
M. Janson\inst{\ref{ins:unisto}},
M. Langlois\inst{\ref{ins:lyon}}
A.~F. Lanza\inst{\ref{ins:oact}},
C. Lazzoni\inst{\ref{ins:exe}}, 
N. Lodieu\inst{\ref{ins:iac}},
A. Maggio\inst{\ref{ins:oapa}}, 
L. Mancini\inst{\ref{ins:unirm2},\ref{ins:oato},\ref{ins:mpia}}, 
E. Molinari\inst{\ref{ins:oaca}},
M. Molinaro\inst{\ref{ins:oats}},
F. Murgas\inst{\ref{ins:iac},\ref{ins:unila}},
D. Nardiello\inst{\ref{ins:oapd},\ref{ins:unimar}}, 
}

\institute{
INAF -- Osservatorio Astrofisico di Torino, via Osservatorio 20, I-10025, Pino Torinese, Italy \label{ins:oato}
\and ICSC -- National Research Centre for High Performance Computing, Big Data and Quantum Computing, Via Magnanelli 2, I-40033, Casalecchio di Reno, Italy \label{ins:icsc}
\and INAF -- Istituto di Astrofisica e Planetologia Spaziali, via Fosso del Cavaliere 100, I-00133, Roma, Italy \label{ins:iaps}
\and Dipartimento di Fisica e Astronomia -- Universit\`a di Padova, Via Marzolo 8, I-35121 Padova, Italy\label{ins:dfapd}
\and INAF -- Osservatorio Astronomico di Padova, Vicolo dell'Osservatorio 5, I-35122, Padova, Italy   \label{ins:oapd}
\and INAF - Osservatorio Astronomico di Trieste, Via G.~B. Tiepolo 11, I-34143, Trieste, Italy\label{ins:oats}
\and  INAF -- Osservatorio Astronomico di Palermo, Piazza del Parlamento, 1, 90134, Palermo, Italy \label{ins:oapa}
\and Space Science Data Center - ASI, Via del Politecnico snc, 00133 Roma, Italy \label{ins:ssdc}
\and Instituto de Astrof{\'\i}sica de Canarias, E-38200 La Laguna, Tenerife, Spain \label{ins:iac}
\and Universidad de La Laguna, Dept. Astrof{\'\i}sica, E-38206 La Laguna, Tenerife, Spain \label{ins:unila}
\and INAF -- Osservatorio Astronomico di Roma,  Via Frascati 33, I-00040, Monte Porzio Catone, Italy \label{ins:oarm}
\and School of Physical Sciences, The Open University, Walton Hall, Milton Keynes MK7 6AA, UK \label{ins:open}
\and SUPA, Institute for Astronomy, University of Edinburgh, Blackford Hill, Edinburgh EH9 3HJ, UK \label{ins:supa}
\and INAF -- Osservatorio Astronomico di Brera, Via E. Bianchi 46, I-23087, Merate, Italy  \label{ins:oabr}
\and Instituto Nacional de Astrof{\'\i}sica, {\'O}ptica y Electr{\'o}nica, Luis Enrique Erro 1, Sta. Ma. Tonantzintla, Puebla, Mexico \label{ins:unimx}
\and Universit\`e C$\hat o$te d'Azur, Observatoire de la C$\hat o$te d'Azur, CNRS, Laboratoire Lagrange, Nice, France \label{ins:cote}
\and Universit\`e Grenoble Alpes, CNRS, IPAG, 38000 Grenoble, France \label{ins:ipag}
\and Geneva Observatory, University of Geneva, Chemin Pegasi 51, 1290 Versoix, Switzerland \label{ins:unige}
\and Department of Astronomy, Stockholm University, AlbaNova University Center, 10691 Stockholm, Sweden \label{ins:unisto}
\and CRAL, UMR 5574, CNRS, Université de Lyon, Ecole Normale Supérieure de Lyon, 46 allée d’Italie, 69364 Lyon Cedex 07, France \label{ins:lyon}
\and INAF -- Osservatorio Astrofisico di Catania, Via Santa Sofia 78, 95123 Catania, Italy \label{ins:oact}
\and Department of Physics and Astronomy, University of Exeter, Stocker Road, Exeter EX4 4QL, UK
 \label{ins:exe}
\and INAF -- Osservatorio Astronomico di Cagliari, Via della Scienza 5, I-09047 Selargius, Italy \label{ins:oaca}
\and Department of Physics, University of Rome ``Tor Vergata'', Via della Ricerca Scientifica 1,
I-00133, Roma, Italy \label{ins:unirm2}
\and Max Planck Institute for Astronomy, K\"{o}nigstuhl 17, D-69117, Heidelberg, Germany \label{ins:mpia}
\and Aix Marseille Univ, CNRS, CNES, LAM, Marseille, France \label{ins:unimar}
}

\date{Received ; accepted }
%\date{2021- January 01  $\rightarrow$ \today}

%\begin{document}
%\maketitle

%\tableofcontents

%\listoffigures

%\listoftables

%\newpage
%\begin{abstract}
%Your abstract.
%\end{abstract}

\abstract{Observational data from space and ground-based campaigns reveal that the 10-30 Ma old V1298\,Tau star hosts a compact and massive system of four planets. \textcolor{black}{Mass estimates are available for the two outer giant planets and point to} unexpectedly high densities for their young ages.}{We investigate the formation of these two outermost giant planets, V1298\,Tau\,b and e, and the present dynamical state of V1298\,Tau's global architecture to shed light on the history of this young and peculiar extrasolar system.}{We perform detailed N-body simulations to explore the link between the densities of V1298\,Tau\,b and e and their migration and accretion of planetesimals within the native circumstellar disk. We combine N-body simulations and the \textcolor{black}{normalized angular momentum deficit (NAMD) analysis of the architecture} to characterize V1298\,Tau's dynamical state and connect it to the formation history of the system. We search for outer planetary companions to constrain V1298\,Tau's planetary architecture and the extension of its primordial circumstellar disk.}{The high densities of V1298\,Tau\,b and e suggest they formed quite distant from their host star, likely beyond the CO$_2$ snowline. The higher nominal density of V1298\,Tau\,e suggests it formed farther out than V1298\,Tau\,b. The current architecture of V1298\,Tau is not characterized by resonant chains. Planet-planet scattering with an outer giant planet is the most likely cause for the instability, but our search for outer companions using SPHERE and GAIA observations can currently exclude only the presence of planets more massive than 2 M$_\textrm{J}$.}{The most plausible scenario for V1298\,Tau's formation is that the system is formed by convergent migration and resonant trapping of planets born in a compact and plausibly massive disk. The migration of V1298\,Tau\,b and e leaves in its wake a dynamically excited protoplanetary disk and naturally creates the conditions for the later breaking of the resonant chain by planet-planet scattering.} 
% 5 {} token are mandatory

\keywords{%Exoplanets -- 
Stars: planetary systems, individual: V1298\,Tau -- Planets and satellites: formation, dynamical evolution and stability, detection -- Chaos}

\titlerunning{The unusual formation history of V1298\,Tau}
\authorrunning{Turrini et al.}

\maketitle

\section{Introduction}\label{sec:introduction}

Young planets offer us the unique opportunity to study %the 
unaltered products of planet formation, before secular evolution and the interactions with their host stars modify or cancel their original characteristics. When they belong to multi-planet systems, they also represent invaluable case studies to comparatively investigate how different formation histories can shape planets born \textcolor{black}{from the same stellar and disk environments.} %around the same star and within the same circumstellar disk. %\citep{turrini2018,turrini2021b}. 

Multiple authors \citep[e.g.][]{donatietal2016nature,yu2017,benatti2021} argue that the population of massive planets in close orbits around young stars can be significantly larger than that of their counterparts around older stars. The proposed roots of this difference are the mass loss in the early evolutionary stages driven by the strong X and UV radiation from the host star, as well as the processes of orbital migration and chaotic evolution at play during the early phases of planetary systems that favour the engulfment \citep[e.g.][]{donatietal2017mnras,spina2021} or the removal \citep[e.g.][]{zinzi2017,turrini2018,Turrini2020,turrini2022} of planets. 

Because young stars are usually very active, another reason for this difference could reside in the false positives induced by stellar activity. The %higher 
number of young planets is debated by recent works that unveil fake detections %of them 
\citep[e.g.][]{carleoetal2018aa, donatietal2020mnras, damassoetal2020aa}.
In the past years, NASA \textit{Kepler} and its K2 campaigns \citep{boruckietal2003spie, howelletal2014pasp}, and recently TESS \citep[Transiting Exoplanet Survey Satellite, ][]{rickeretal2015jatis} detected transits of several young and multiple planets \citep[e.g.][]{davidetal2016apj816_21, mannetal2016aj, rizzuto2020aj, plavchanetal2020nature}, allowing the study of the early evolution and architecture of these systems \citep[e.g.][]{benattietal2019aa, benattietal2021aa650_66}.

\textcolor{black}{Analyzing the \textit{Kepler} light curve of the young star V1298\,Tau  \citep[K2 Mission, ][]{howelletal2014pasp126_398}, \citet{davidetal2019aj158_79} discovered a first Jupiter-sized planet (V1298\,Tau\,b) to which \citet{davidetal2019apjl885_l12} soon added other three planets after a forward analysis of the K2 Campaign 4 photometry. The transits of all four planets have been later observed with TESS \citep{feinsteinetal2022apjl925_2}, confirming that V1298 Tau's system is host to two Neptune-sized planets (dubbed ``c'' and ``d'') and two Jovian-sized planets (``b'' and ``e''), in order of distance from the parent star. The four planets inhabit an orbital region comparable to that contained within Mercury's orbit in the Solar System.}

\textcolor{black}{V1298\,Tau has later been observed, in collaboration with other groups, in the framework of the GAPS project \citep[Global Architecture of Planetary Systems,][]{covinoetal2013aa,carleo2020} by means of an intensive spectroscopic campaign. The GAPS campaign attained radial velocity (RV) measurements using several high-resolution spectrographs \citep[HARPS-N, CARMENES, SES and HERMES;][]{suarezmascarenoetal2021natast6_232} to constrain the masses of the four planets. These  observations \citep{suarezmascarenoetal2021natast6_232}, combined with the constraints from \textit{Kepler} and K2, yield mass values for both planet b ($0.64 \pm 0.19$ M$_{\text{J}}$) and planet e ($1.16 \pm 0.30$ M$_{\text{J}}$), while for planets c and d they provide upper limits only ($<$0.24 and $<$0.31 M$_\text{J}$ respectively).} \textcolor{black}{Notwithstanding the uncertainties affecting the mass estimates, the GAPS survey reveals that V1298\,Tau is one of the most massive systems among those characterized by compact orbital architectures discovered so far.} 

\textcolor{black}{Using the \texttt{ExoplAn3T} online tool (Exoplanet Analysis and 3D visualization Tool\footnote{\url{https://tools.ssdc.asi.it/exoplanet/}}, \citealt{zinzi2021b,zinzi2021a}) developed by the Space Science Data Center of the Italian Space Agency, we queried the NASA Exoplanet Archive\footnote{\url{https://exoplanetarchive.ipac.caltech.edu/}} on 8th May 2023 to search for massive exoplanetary systems with similarly compact  architectures. We searched for systems with multiplicity greater than two around solar-type stars, orbital periods lower than 100 days, and hosting at least one planet with mass greater than 100 M$_\oplus$. The only systems satisfying these constraints are V1298\,Tau, Kepler-46 (three planets, two giants) and Kepler-256 (four planets, one giant). Adding the further requirement that the total planetary mass is greater than 1 M$_\textrm{J}$ restricts this list to V1298\,Tau and Kepler-46 only.}

\textcolor{black}{The availability of both planetary radii and masses for planets b and e allows for the  estimation of their bulk densities. The values derived by \citet{suarezmascarenoetal2021natast6_232} using the planetary radii from \textit{Kepler} and K2 are $1.2\pm0.45$ and $3.6\pm1.6$\,g\,cm$^{-3}$ for planets b and e, respectively. While affected by large uncertainties, these values are significantly greater than those predicted by formation theories for their age.} Two possible explanations are discussed by \citet{suarezmascarenoetal2021natast6_232}: the more rapid contraction of these young planets than predicted by interior evolution models, or their extreme enrichment in heavy elements. 

While the first possibility implies the need to revise our current understanding of giant planet evolution \citep{suarezmascarenoetal2021natast6_232}, the second explanation naturally arises from the planet formation process when giant planets undergo extensive migration within their native disk \citep{Thorngren2016,Shibata2020,Turrini2021}. \textcolor{black}{The conclusions drawn by \cite{suarezmascarenoetal2021natast6_232} are valid also for the new measurements of the radii of V1298\,Tau's planets by the photometric observations from TESS \citep{feinsteinetal2022apjl925_2}, which update %have revised 
upward and downward the densities of planets b and e, respectively (roughly by about 30\%).}

The architecture of V1298\,Tau's planets \textcolor{black}{initially appeared} very close to being in a resonant chain. In particular, the orbit of the fourth planet was originally assessed by \cite{suarezmascarenoetal2021natast6_232} to be close to completing a resonant chain, either 3:2, 2:1, 3:2 or 3:2, 2:1, 2:1.
\textcolor{black}{However, the period ratio between planet b and d is too far from a 2:1 resonant ratio\, effectively excluding the possibility that the system is currently in a resonant chain \citep{Tejada2022}.} 
%In addition  planet e was observed in transit only twice (with Kepler and TESS), hence its period is yet to be conclusively quantified. 
Furthermore, recent data indicate an orbital period of planet e \citep{feinsteinetal2022apjl925_2, Damasso2023} that is not in resonance with that of planet b.
%and favour V1298\,Tau's current architecture being highly unstable \citep{Tejada2022}.} 
%

The combination of the \textcolor{black}{masses and} densities of planets b and e and the compact architecture of V1298\,Tau points toward the system having formed by convergent migration within the circumstellar disk followed by resonant trapping. \textcolor{black}{The trapping in a stable resonant configuration is particularly important for planets b and e, which otherwise should go unstable long before reaching their current compact orbits. In this scenario, however, the current non-resonant orbits of its planets require that the system underwent a phase of instability that broke the original resonant chain notwithstanding its young age \citep{Tejada2022}.}
%Furthermore, while V1298\,Tau's compact architecture can be dynamically stable in the case of low--eccentricity orbits \citep{Tejada2022}, the results from the GAPS campaign point to markedly eccentric orbits  for both planets b and e \citep{suarezmascarenoetal2021natast6_232}. 
%Therefore, the big question is why V1298\,Tau's compact architecture is present regardless of not being characterized by resonant-locking.}% and plausibly having undergone a phase of dynamical instability.}
%nor by low--eccentricity orbits}. 

In this work, %we investigate this apparent contradiction by jointly studying 
%jointly study the formation of V1298\,Tau's b and e and the stability of the global architecture of the planetary system, 
\textcolor{black}{we jointly study the formation of V1298\,Tau's b and e and of their orbital architecture with the aim to shed light on V1298\,Tau's unusual formation history}. We complement our study with new observations of V1298\,Tau, searching for additional outer planets to constrain the extension of the planet-forming region in V1298\,Tau's circumstellar disk. As we will show, the convergent and large-scale migration required to explain the high density values and compact orbits of V1298\,Tau\,b and e naturally creates the conditions to evolve the planetary system into its present configuration.

This paper is organized as follows: \textcolor{black}{in Sects. \ref{sec:v1298tau} and \ref{sec:densities} we revise the fundamental parameters of V1298 Tau's planets based on the most up-to-date observations, while in Sect. \ref{sec:methods} we} describe the numerical algorithms used in modelling \textcolor{black}{their formation}, their capture in resonance and the onset of their subsequent instability. % as well as to characterize the stability of the system. 
In Sect. \ref{sec:results} we outline our results concerning the %\textcolor{black}{the dynamical state and} 
history of the planetary system and the conditions that can lead to its present 
%unstable %dynamically excited
\textcolor{black}{non-resonant} configuration. Sect. \ref{sec:observations} is devoted to observational constraints on the presence of additional planets on outer orbits obtained with SPHERE (the Spectro-Polarimetric High-contrast Exoplanet REsearch facility at the VLT telescope), while in Sect. \ref{sec:conclusions} we discuss the implications of our results and combine them in a unified picture.

\section{V1298\,Tau: the star and its planets}\label{sec:v1298tau}

\begin{figure*}[t]
    \centering
    \includegraphics[width=\textwidth]{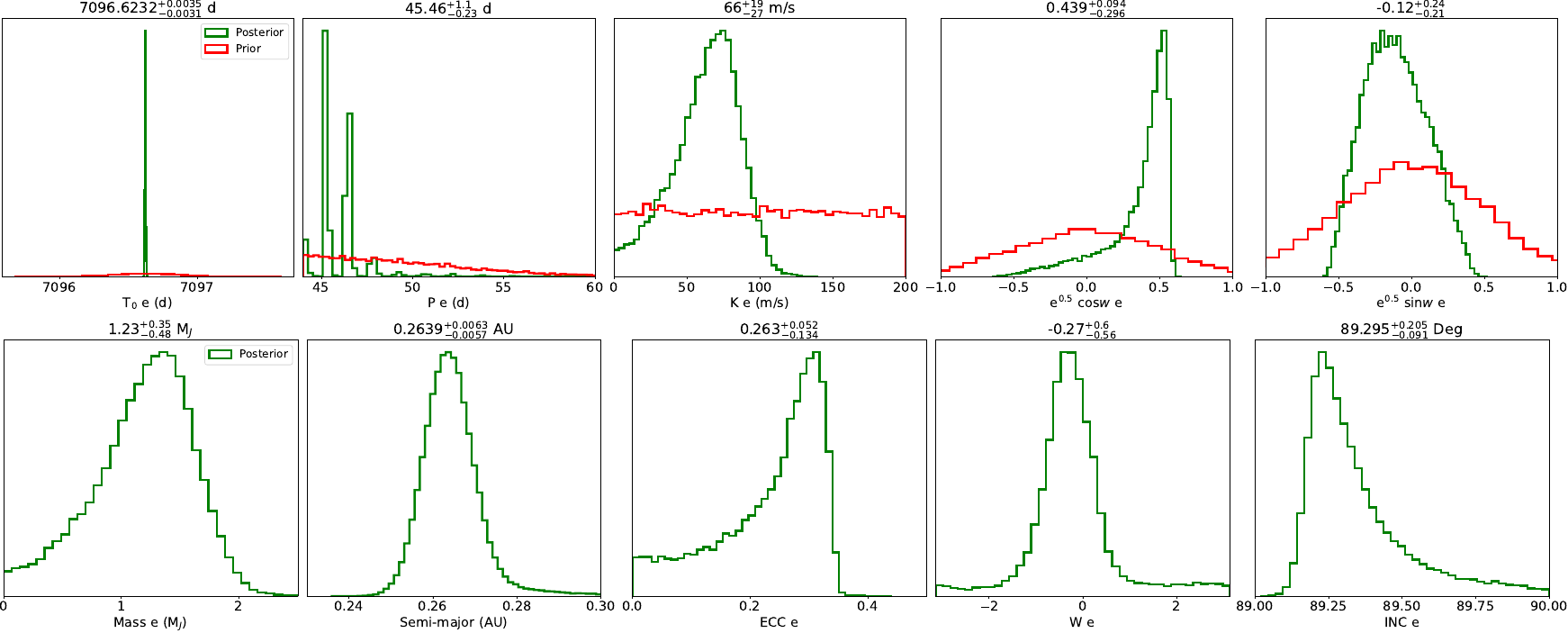}
    \caption{\textcolor{black}{New analysis of the RV dataset and its results on planet e. \textit{Top}: priors and posteriors of the new analysis from \cite{suarezmascarenoetal2021natast6_232} where we include the constraints from TESS on its orbital period \citep{feinsteinetal2022apjl925_2}. \textit{Bottom}: derived distributions of the planetary and orbital parameters. Note that we use the updated planetary radius from TESS \citep{feinsteinetal2022apjl925_2} in our analysis in place of the one appearing in this figure, which is still based on the \textit{Kepler} and K2 data as in \cite{suarezmascarenoetal2021natast6_232}.}}
    \label{fig:newretrieval}
\end{figure*}

\textcolor{black}{V1298\,Tau, with its estimated age ranging between 10 and 30 Ma \citep{suarezmascarenoetal2021natast6_232,maggioetal2022apj925_172}, is one of the youngest solar-type planet-host stars. %The stellar mass and radius are $1.170 \pm 0.060$\ M$_\odot$, and $1.278 \pm 0.070$\ R$_\odot$, respectively. 
With its $T_{\rm eff} = 5050 \pm 100$ K, V1298\,Tau is a K1 spectral type with iron abundance $[{\rm Fe/H}]=0.10 \pm 0.15 $\,dex \citep{suarezmascarenoetal2021natast6_232} belonging to the young stellar group 29 identified by \citet{ohetal2017aj153_257O}. V1298\,Tau is characterized by a high and steady activity level, causing large RV activity variations, as testified also by the value of $<\log (\text{R}'_{\text{HK}})>$=$-4.2397 \pm 0.0179$ determined by GAPS \citep{suarezmascarenoetal2021natast6_232}. We report the main stellar parameters of V1298\,Tau in Table\ \ref{tab:v1298par}.}
%
%.  V1298\,Tau is physically tied  to the G2 star HD\,284154, and both stars belong to the young stellar group 29 identified by \citet{ohetal2017aj153_257O}. 
%

\textcolor{black}{From XMM-Newton observations, \citet{maggioetal2022apj925_172} found that this young star has a bolometric luminosity ratio of $\log \text{L}_{\text{X}}/\log \text{L}_{\text{Bol}}=-3.35^{+0.01}_{-0.02}$ confirming that it is an X-ray bright young star near the saturated emission regime observed for G-K stars. Despite the close position of the four planets to their star (see Table \ref{tab:v1298par}), \citet{maggioetal2022apj925_172} found that the two outer planets (b and e) are not affected by evaporation on Gyr-long timescales. The two inner planets (c and d) are also impervious to evaporation if their masses are higher than $40\ \text{M}_\oplus$.}

%, while we refer readers to Table 1 of \citet{suarezmascarenoetal2021natast6_232} for a complete description of the stellar companion of V1298\,Tau. 

%\textcolor{black}{V1298 Tau's system is host to} two Neptune-sized planets (dubbed ``c'' and ``d'') and two Jovian planets (``b'' and ``e''), in order of distance from the central star. 
%The observations by the GAPS campaign \citep{suarezmascarenoetal2021natast6_232}\textcolor{black}{, combined with the constraints from the \textit{Kepler} and K2 observations, yield mass values for} both planet b ($0.64 \pm 0.05$ M$_{\text{J}}$) and planet e ($1.16 \pm 0.30$ M$_{\text{J}}$), while for planets c and d they can provide upper limits only ($<$0.24 and $<$0.31 M$_\text{J}$ respectively). %, which however were affected by V1298\,Tau's activity. 

\textcolor{black}{The analysis performed by \cite{suarezmascarenoetal2021natast6_232} to derive the mass of planet e attributed an orbital period of 40 days to the planet. The subsequent observations by TESS revised this value upward, suggesting a most probable period of 44 days and providing additional discreet solutions for larger orbital periods characterized by decreasing probabilities \citep{feinsteinetal2022apjl925_2}. The observations by TESS also revised downward and upward the planetary radii of planets b and e, respectively \citep{feinsteinetal2022apjl925_2}. Following these new observations, we reevaluate the mass of planet e and then update its density and that of planet b.}

\textcolor{black}{To reassess the mass of planet e accounting for the larger possible values of its orbital period, we reanalyse the RV dataset from \cite{suarezmascarenoetal2021natast6_232} imposing the additional constraint to the priors that the orbital period of planet e should not be smaller than 44 days. \cite{feinsteinetal2022apjl925_2} show that a orbital period shorter than this value would result in an unobserved second transit within the TESS baseline. Previously, K2 data only constrained the orbital period to values higher than 36 days. Following the results of \cite{feinsteinetal2022apjl925_2} we used a half-normal distribution starting at 44 days, with a sigma of 7 days, as prior for the period of planet e. Aside for this additional prior, the analysis is the same as described in \cite{suarezmascarenoetal2021natast6_232}}. 

\textcolor{black}{We fit K2 photometry, ground-based photometry and RV time-series simultaneously. We model the activity signals in all datasets using Gaussian Processes regression with \texttt{celerite}~\citep{Foreman-Mackey2017}. The K2 photometry is used to obtain information on the parameters of the transits. The ground-based photometric data is contemporary to the RV data and helps constrain the parameters of the stellar activity model. To sample the posterior distribution we rely on Nested Sampling \citep{Skilling2004} using \texttt{dynesty} \citep{Speagle2020}. For more details and a full description of the priors used in the global model we refer the readers to \cite{suarezmascarenoetal2021natast6_232}.}

\textcolor{black}{The priors and posteriors of the analysis are shown in the top row of Fig. \ref{fig:newretrieval}, while the bottom row shows the derived planetary and orbital parameters of planet e. The new mass value of planet e is 1.23$^{+0.35}_{-0.48}$ M$_\text{J}$, which is slightly higher yet consistent within 1-$\sigma$ with the previous value from \cite{suarezmascarenoetal2021natast6_232}.} \textcolor{black}{The analysis also updates the plausible periods of V1298\,Tau\,e and shows that the most likely values are at 45 and 46 days, with two possible lower-probability solutions at 44 and 47.7 days (see Fig. \ref{fig:newretrieval}).} \textcolor{black}{While multi-modal posteriors like those characterizing the orbital period are known to be difficult to sample with traditional MCMC methods, nested sampling methods in general, and \texttt{dynesty} in particular, have been shown to be able to sample them robustly and efficiently \citep{Speagle2020}.} 

None of the solutions presented above results in a resonant coupling between planets b and e. Preliminary results from ongoing observations of the ESA mission CHEOPS exclude the solutions at 44 and 46 days, and are instead compatible with the one at about 45 days \citep{Damasso2023}, which we adopt as the nominal period of V1298\,Tau\,e of this work. The full updated characterization of V1298\,Tau's four planets is reported in Table \ref{tab:v1298par}. Fig. \ref{fig:newretrieval} and Table \ref{tab:v1298par} show that, like its mass, the eccentricity of planet e is higher than that estimated by \cite{suarezmascarenoetal2021natast6_232} but also in this case the two values are consistent at 1-$\sigma$ level. %As discussed in \citet{suarezmascarenoetal2021natast6_232}, the orbital parameters of the four planets are constrained through the analysis of the RV time series and are therefore affected by V1298\,Tau's high activity. 

% Modal NAMD value plus lower and upper boundary: 2.2e-2, 6.4e-3, 0.157 -> 17x, 5x,  120x w.r.t. Solar System (the lower and upper boundaries provide the 3-sigma confidence interval)

%V1298\,Tau is characterized by a high and steady activity level, causing large RV activity variations, as testified also by the value of $<\log (\text{R}'_{\text{HK}})>$=$-4.2397 \pm 0.0179$ determined by GAPS. From XMM-Newton observations, \citet{maggioetal2022apj925_172} found that this young star has a bolometric luminosity ratio of $\log \text{L}_{\text{X}}/\log \text{L}_{\text{Bol}}=-3.35^{+0.01}_{-0.02}$ confirming that it is an X-ray bright young star near the saturated emission regime observed for G-K stars. Despite the close position of the four planets to their star, \citet{maggioetal2022apj925_172} found that the two outer planets (b and e) are not affected by evaporation on Gyr-long timescales. The two inner planets (c and d) are also impervious to evaporation if their masses are higher than $40\ \text{M}_\oplus$.

%Before investigating the formation histories of V1298\,Tau\,b and e, we %update their estimated densities by combining their mass values from \citet{suarezmascarenoetal2021natast6_232} with the new planetary radii from TESS by \citet{feinsteinetal2022apjl925_2}

%\textcolor{black}{Following the new observational constraints from TESS \citep{feinsteinetal2022apjl925_2}, we revise the characterization of the mass of planet e before updating the estimated densities of both planet b and e.}
\begin{table}[t]
    \centering
    \begin{tabular}{lcc}
    \hline
Parameters & Value  & Reference\\
\hline
\hline
\multicolumn{3}{l}{\textbf{Stellar Parameters}}\\
R$_\star$ (R$_\odot$)      & $1.33^{+0.04}_{-0.03}$& (1*)\\
M$_\star$ (M$_\odot$)      & $1.17 \pm 0.06$      & (2)\\
B (mag)                    & $11.11 \pm 0.09$     & (3)\\
V (mag)                    & $10.12 \pm 0.05$     & (3)\\
G (mag)                    & $10.0702 \pm 0.0007$ & (4)\\
J (mag)                    & $8.687 \pm 0.023$    & (5)\\
Spectral Type              & K1                   & (6)\\
T$_{\text{eff}}$(K)        & $5050 \pm 100$       & (2)\\
Parallax (mas)             & $9.258 \pm 0.020$    & (4)\\
Distance (pc)              & $108.5 \pm 0.7$      & (4)\\
%$<\log (\text{R}'_{\text{HK}})>$& $-4.2397 \pm 0.0179$ &(6)\\
$<\log (\text{R}'_{\text{HK}})>$& $-4.240 \pm 0.018$ &(7)\\
Age (Ma)                  & $10-30$              & (2,8)\\
% $ 11.9^{+2.0}_{-3.2}$& (7)\\
\hline
\multicolumn{3}{l}{\textbf{Planetary Parameters}}\\
\multicolumn{3}{l}{\textbf{Planet b}}\\
R$_p$/R$_*$                 & $0.0636\pm0.0018$  & (1)\\
R$_\text{P}$ (R$_\text{J}$) & $0.84 \pm 0.02$   & (7)\\
M$_\text{P}$ (M$_\text{J}$) & $0.64 \pm 0.19$   & (2)\\
P$_\text{Orb}$ (d)          & $24.1315^{+0.00033}_{-0.0034}$ & (1)\\
a (au)                      & $0.1719 \pm 0.0027$   & (2)\\
inclination ($^\circ$)      & $>88.7$  & (2)\\
eccentricity                & $0.134 \pm 0.075$   & (2)\\
\multicolumn{3}{l}{\textbf{Planet c}}\\
R$_p$/R$_*$                 & $0.0337\pm0.0009$  & (1)\\
R$_\text{P}$ (R$_\text{J}$) & $0.45 \pm 0.01 $ & (7)\\
M$_\text{P}$ (M$_\text{J}$) & $ <0.24 $   & (2)\\
P$_\text{Orb}$ (d)          & $ 8.2438^{+0.0024}_{-0.0020}$   & (1)\\
a (au)                      & $ 0.0841 \pm 0.0013$   & (2)\\
inclination ($^\circ$)      & $>87.5 $  & (2)\\
eccentricity                & $<0.30$  & (2)\\
\multicolumn{3}{l}{\textbf{Planet d}}\\
R$_p$/R$_*$                 & $0.0409^{+0.0014}_{-0.0015}$  & (1)\\
R$_\text{P}$ (R$_\text{J}$) &  $0.54 \pm 0.02$   & (7)\\
M$_\text{P}$ (M$_\text{J}$) & $ <0.31 $   & (2)\\
P$_\text{Orb}$ (d)          &  $ 12.3960^{+0.0019}_{-0.0020}$  & (1)\\
a (au)                      &  $ 0.1103 \pm 0.0017$  & (2)\\
inclination ($^\circ$)      &  $>88.3$  & (2)\\
eccentricity                &  $<0.20$ & (2)\\
\multicolumn{3}{l}{\textbf{Planet e}}\\
R$_p$/R$_*$                 & $0.0664^{+0.0025}_{-0.0021}$  & (1)\\
R$_\text{P}$ (R$_\text{J}$) & $0.88 \pm0.03$   & (7) \\
M$_\text{P}$ (M$_\text{J}$) & 1.23$^{+0.35}_{-0.48}$ & (7)\\%& $1.16 \pm 0.30$   &(1)\\
P$_\text{Orb}$ (d)          & 45.46$^{+1.10}_{-0.23}$ & (7)\\%&  $44.1699$  & (8)\\
a (au)                      & 0.2639$^{+0.0062}_{-0.0057}$ & (7)\\%&  $0.2409 \pm 0.0083$  & (1)\\
inclination ($^\circ$)      & 89.295$^{+0.205}_{-0.091}$ & (7)\\%&  $>89.0$  & (1)\\
eccentricity                & 0.263$^{+0.052}_{-0.134}$ & (7)\\%&  $ 0.10 \pm 0.091$  & (1)\\
\hline
\end{tabular}
\caption{Main stellar and planetary parameters of V1298\,Tau's system.}
\label{tab:v1298par}
\tablebib{(1) \citet{feinsteinetal2022apjl925_2}; (2) \citet{suarezmascarenoetal2021natast6_232}; (3) \citet{hogetal2000aa355_27}; (4) \citet{gaiacoll2021aa649_1}; (5) \citet{skrutskieetal2006aj131_1163}; (6) \citet{nguyenetal2012apj745_119}; (7) This work; (8) \citet{maggioetal2022apj925_172}; * \cite{suarezmascarenoetal2021natast6_232} report a slightly smaller value of R$_*$=$1.29\pm0.07$. As the two values are consistent with each other within 1 $\sigma$, we adopt the value of \cite{feinsteinetal2022apjl925_2} for consistency with their fitting of TESS light-curves.}
\end{table}

\subsection{Dynamical excitation of V1298\,Tau b and e}\label{sec:namd}

\textcolor{black}{We use the updated architecture of V1298\,Tau from Table \ref{tab:v1298par} to investigate the dynamical excitation of planets b and e} by means of their normalized angular momentum deficit \citep[NAMD, see][]{chambers2001,Turrini2020}. The NAMD provides an architecture-agnostic measure of dynamical excitation that can intuitively be interpreted as the ``dynamical temperature'' of planetary systems: the higher the value, the more excited is the dynamical state of the system. We adopt the NAMD of the Solar System ($1.3\times10^{-3}$, \citealt{Turrini2020}) as the boundary between dynamically cold and hot orbits (see \citealt{carleo2021,turrini2022} for a discussion).

\textcolor{black}{Following \citet{carleo2021}, the goal of our analysis is not to pinpoint the exact value of V1298\,Tau's NAMD but to verify if, notwithstanding the uncertainties \textcolor{black}{on the physical and orbital parameters of its planets}, its dynamical excitation is systematically higher than that of the Solar System and indicates past or current phases of dynamical instability. We account for the uncertainties on the mass, semimajor axis and eccentricity of the two planets following the Monte Carlo approach described in \cite{Turrini2020} and \cite{carleo2021}. To better sample the effects of the large uncertainties of these parameters on the NAMD, we use 10$^{6}$ extractions for each parameter in place of the original 10$^{4}$ proposed by \citet{LaskarPetit2017}.}

\textcolor{black}{Based on the low mutual inclination of the two planets (<1$^{\circ}$, see Table \ref{tab:v1298par}) that limitedly contributes to the NAMD, we assume the two orbits as coplanar. Due to the positive-defined nature of the NAMD, this choice means that we err toward lower excitations. The resulting NAMD modal value ($2.2\times10^{-2}$) is 17 times higher than that of the Solar System, with the 3$\sigma$ range extending from 5 times ($6.4\times10^{-3}$) to 120 times ($0.16$) that of the Solar System. As discussed by \citet{Turrini2020,turrini2022} and illustrated by the recent study by \citet{rickman2023}, such a systematically high range of NAMD values is associated to phases of dynamical instability and planet-planet scattering events (either in the past or presently).}

\textcolor{black}{The results of the NAMD analysis quantitatively confirm the dynamical indication supplied by the present non-resonant state of this compact system \citep{Tejada2022}. Specifically, the dynamical state of V1298\,Tau  argues in favour of a formation scenario where the planetary system acquired its compact architecture by forming by convergent migration and resonant capture. The original resonant architecture was broken during a later phase of dynamical instability. This scenario is supported by dynamical population studies of exoplanetary systems \citep{LaskarPetit2017,gajdos2023} revealing how 25-45\% of known multi-planet systems, and the majority of high-multiplicity systems hosting four or more planets like V1298\,Tau, show the signatures of chaos and instability in their architectures. We investigate the plausible causes of the instability in Sects. \ref{sec:planet-planet_scattering} and \ref{sec:planetesimal_perturbations}.}% even in presence of collisions or loss of planets.}

\subsection{Density values of V1298\,Tau\,b and e}\label{sec:densities}

\textcolor{black}{We reevaluate the densities of planets b and e using the new mass estimate of planet e from Sect. \ref{sec:v1298tau} and the planetary radii provided by TESS \citep{feinsteinetal2022apjl925_2}. Specifically, we use} the values of $R_{\rm p}/R_*$ and $R_*$ from \cite{feinsteinetal2022apjl925_2} reported in Table \ref{tab:v1298par} together with the volumetric mean radii of the Sun and Jupiter\footnote{\url{https://nssdc.gsfc.nasa.gov/planetary/planetfact.html}} to estimate the planetary radii (see Table \ref{tab:v1298par}) and volumes. The resulting density values are $1.4\pm0.4$\,g\,cm$^{-3}$ and \textcolor{black}{$2.4\pm0.8$\,g\,cm$^{-3}$} respectively. As discussed by \citet{suarezmascarenoetal2021natast6_232}, unless the two giant planets underwent %significantly
a more rapid contraction than predicted by interior evolution models, their densities %are 
can be explained by marked enrichments in heavy elements \citep{Thorngren2016}. 

%Turrini2019,tychoniec2020,
\textcolor{black}{Recent results on the evolution over time of the dust abundance in circumstellar disks \citep{Manara2018,mulders2021,bernabo2022}} and the radiometric ages of meteorites in the Solar System \citep[see][and references therein]{scott2007,coradini2011,lichtenberg2022} argue that the bulk of the heavy elements is locked into planetesimals by the time giant planets form. As discussed by \citet{Shibata2020} and \citet{Turrini2021} the mass of planetesimals that giant planets accrete during their growth is directly proportional to the extent of their migration. The larger the migration, the greater the mass of planetesimals that can enter their feeding zone and be accreted. Conversely, giant planets undergoing little or no migration will experience limited accretion of planetesimals \citep{turrini2015,shibata2019}. 
%The marked enrichments in heavy elements of V1298\,Tau\,b and e therefore \textcolor{black}{suggest their extensive migration}. %The higher density of planet e suggests that it formed farther out from the star than planet b.

We use the updated density values of the two giant planets to derive order-of-magnitude estimates of the masses of planetesimals that they need to accrete. We focus on the density of the solid material and not on the bulk density of the planetesimals (as instead we do in the N--body simulations) to remove the issue of the unknown macroporosity of the planetesimals. We assume the solid material accreted through the planetesimals to possess density of 3\,g\,cm$^{-3}$, i.e. to be composed half of rock and metals with average density of 5\,g\,cm$^{-3}$ and half of ice with density of 1\,g\,cm$^{-3}$ \citep[see e.g.][for the mass balance between rock and ice]{Turrini2021,pacetti2022}. 

\textcolor{black}{For the gas composing the envelopes of V1298\,Tau\,b and e we assume} the same metallicity and density as that of Jupiter in the Solar System. This gas has density of 1.33\,g\,cm$^{-3}$ and \textcolor{black}{its composition is about three times richer in heavy elements with respect to hydrogen than that of the Sun \citep[see][and references therein]{atreya2018,oberg2019}, i.e. heavy elements account for at least 4\% of Jupiter's mass.} 
%the solar-metallicity gas composing V1298\,Tau and its circumstellar disk \citep{suarezmascarenoetal2021natast6_232}. 
\textcolor{black}{As in the enrichment scenario discussed by \cite{suarezmascarenoetal2021natast6_232} for planets b and e, Jupiter's formation, metallicity and enrichment in heavy elements are argued to have been shaped by large-scale migration and planetesimal accretion \citep{pirani2019,oberg2019}.}%,shibata2022}.}
%For simplicity, following \cite{oberg2019} and \cite{bosman2019} we assume the gas metallicity to arise from the erosion of the planetary core or the accretion of disk gas enriched in heavy elements.

\textcolor{black}{The density of V1298\,Tau\,b can be obtained by adding about 11\,M$_\oplus$ of planetesimals to about 192\,M$_\oplus$ of Jovian gas. Merging the contributions in heavy elements of planetesimals and enriched Jovian gas, the resulting mixture is composed at 91\% (184\,M$_\oplus$) of H and He and 9\% (19\,M$_\oplus$) of heavy elements.  
The same approach applied to V1298\,Tau\,e requires combining 140 M$_\oplus$ of Jovian gas and 250 M$_\oplus$ of planetesimals. Grouping the heavy elements together results in a mixture where 66\% of the mass (256\,M$_\oplus$) is provided by heavy elements and 34\% (135\,M$_\oplus$) by H and He. %Taken at face value, these back-of-the-envelope estimates would suggest that, 
While V1298\,Tau\,b is similar to Jupiter and Saturn in terms of ratio between heavy elements and hydrogen, V1298\,Tau\,e appears closer to a Neptunian planet as %, while a giant planet,
H and He do not dominate its mass.}

\textcolor{black}{Since this metallicity is anomalously high for such a massive planet \citep{Thorngren2016} \textcolor{black}{and the uncertainty toward lower mass values is particularly large}, we explore the case where the real mass of V1298\,Tau\,e is 0.75 M$_\text{J}$, i.e. 1-$\sigma$ less than the nominal value from Table \ref{tab:v1298par} and Fig. \ref{fig:newretrieval}. This results in a planetary density of 1.5 g\,cm$^{-3}$ and requires adding 29 M$_\oplus$ of heavy elements to 209 M$_\oplus$ of H and He. In this scenario, the metallicities of V1298\,Tau\,b and e are similar and both giant planets are consistent with the ratio between heavy elements and hydrogen of Jupiter and Saturn.}

\textcolor{black}{In the following analysis we will consider both scenarios summarised in Table \ref{tab:scenarios}, one where V1298\,Tau\,e is characterized by higher mass and metallicity (HMZe in the following and in Table \ref{tab:scenarios}) and one where the giant planet has lower mass and metallicity (LMZe in the following and in Table \ref{tab:scenarios}). Before proceeding, we note that the gas density we adopted in our back-of-the-envelope computations is the one characterizing the \textit{present day} Jupiter. The radius of Jupiter between 10-30 Ma after its formation is expected to have been 1.5-1.6 times the present value \citep{Lissauer2009,DAngelo2021}, resulting in its larger volume by a factor 3-4 and correspondingly lower density of the gas (about 0.3-0.4 g\,cm$^{-3}$).}

\textcolor{black}{Such lower gas density requires larger amounts of heavy elements to fit the estimated densities, specifically 87 M$_\oplus$ for V1298\,Tau\,b and 108 M$_\oplus$ for V1298\,Tau\,e even in the LMZe scenario. In this case the two giant planets would be composed by heavy elements for about 45\% of their mass ($Z/Z_*$$\sim$25 using the nominal metallicity of V1298\,Tau from Sect. \ref{sec:v1298tau}). In the HMZe scenario, V1298\,Tau\,e would require about 305 M$_\oplus$ of heavy elements, i.e. H and He would supply only 20\% of its mass ($Z/Z_*$$\sim$45). Given the uncertainty affecting the masses of the two planets and its significant impact on their densities and metallicities (see Table \ref{tab:scenarios}), we do not simulate also these scenarios but we will discuss their implications in Sects. \ref{sec:results-formation} and \ref{sec:conclusions}.}

%about 180 M$_\oplus$ of Jovian gas with about 190 M$_\oplus$ of planetesimals. Given the large uncertainty ($\approx30\%$) on the density values of the two giant planets and on the orbital period of V1298\,Tau\,e, we do not attempt more detailed calculations and only use the estimated values as order-of-magnitude references in our analysis. Within the limits of these back-of-the-envelope calculations, the two giant planets appear to have initially possessed similar masses of gas, their present masses arising from radically different formation and migration tracks.

\begin{table}[t]
    \centering
    \begin{tabular}{ccccc}
    \hline
    \multirow{2}{*}{Scenario Id.} & \multicolumn{2}{c}{V1298\,Tau\,b} & \multicolumn{2}{c}{V1298\,Tau\,e} \\
    & Mass (M$_\text{J}$)& Z (M$_\oplus$) & Mass (M$_\text{J}$)& Z (M$_\oplus$) \\
    \hline
    %Low Z$_e$ 
    \textit{LMZe} & 0.64  & 19 & 0.75 & 29 \\
    %High Z$_e$
    \textit{HMZe} & 0.64  & 19 & 1.23 & 256 \\
    \hline
    \end{tabular}
    \caption{Physical scenarios considered for the planetary masses and the enrichment in heavy elements of V1298\,Tau\,b and V1298\,Tau\,e.}\label{tab:scenarios}
\end{table}

\section{V1298 Tau's formation and dynamical histories: numerical methods}\label{sec:methods}

\subsection{Formation simulations of V1298\,Tau\,b and e}\label{sec:model-formation}

We simulate the growth of V1298\,Tau\,b and e, their migration and interactions with the planetesimal disk with the N-body code \texttt{Mercury-Ar$\chi$es} \citep{Turrini2019,Turrini2021}. 
The N-body simulations model the effects of the mass growth of the forming V1298\,Tau\,b and e planets, their planetary radius evolution and orbital migration, as well as the dynamical evolution of their surrounding planetesimal disk under the effects of their gravitational perturbations alongside those of gas drag and the disk gravity. 

We model the native circumstellar disk as possessing characteristic radii $r_{c}$= 50 AU and gas surface density $\Sigma(r)=\Sigma_{0} \left( r/r_{c} \right)^\gamma exp\left[-\left(r/r_{c} \right)^{\left( 2-\gamma \right)}\right]$ (see Sect. \ref{sec:results-formation} for the values of $\Sigma_{0}$ considered in the simulations) where $\gamma=0.8$ \citep{Isella2016}. The disk gas mass is assumed in steady state and does not decline over time. The disk temperature profile on the midplane is $T(r)= T_{0}\,r^{-0.6}$ where $T_{0}$=200 K \citep{andrews2007,oberg2011,eistrup2016}. The mass of V1298\,Tau is set to 1.17 M$_\odot$ \citep{suarezmascarenoetal2021natast6_232}.

Planetesimals are included in the N-body simulations as particles possessing inertial mass, computed assuming a common diameter of 100 km \citep[see][]{KlahrSchreiber2016,johansen2017,Turrini2019} and bulk density of 1 g/cm$^3$ \citep[see][]{Turrini2019,Turrini2021}, and no gravitational mass. The density of the planetesimals is lower than that used in Sect. \ref{sec:v1298tau} to estimate the amounts of solid material accreted by the giant planets to account for the macro-porosity of these planetary bodies. The dynamical evolution of planetesimals is affected by the gravity of the host star, the forming planets V1298\,Tau\,b and e, and by the disk gas through aerodynamic drag and gravity (as they possess inertial mass). 

The dynamical evolution of the planetesimals is not affected by the interactions among planetesimals themselves (as they do not possess gravitational mass) nor do the planetesimals perturb the two forming planets. The damping effects of gas drag on the planetesimals are simulated following the treatment from \citet{brasser2007} with updated drag coefficients from \citet{nagasawa2019} accounting for both the Mach and Reynolds numbers of the planetesimals. The exciting effects of the disk self-gravity are simulated based on the analytical treatment for axisymmetric disks by \citet{ward1981} following \citet{marzari2018} and \citet{nagasawa2019}.% (see \citealt{Turrini2021} for further discussion).

\textcolor{black}{The formation of the two giant planets is modelled over two growth and migration phases %\citep{Lissauer2009,DAngelo2010,Bitsch2015,johansen2017,johansen2019,DAngelo2021} 
using the parametric approach from \citet{Turrini2019,Turrini2021}. The first phase accounts for their core growth and subsequent capture of an expanded atmosphere \citep[e.g.][]{Bitsch2015,johansen2019,DAngelo2021}. The planetary mass evolves as 
%\begin{equation}
$ M_{p}(t)=M_{0}+\left( \frac{e}{e-1}\right)\left(M_{1}-M_{0}\right)\left( 1-e^{-t/\tau_{p}} \right)$
%\label{eqn-coregrowth}
%\end{equation} 
where M$_{0}$=0.1 M$_{\oplus}$ is the initial mass of the core, M$_{1}$=30 M$_\oplus$ is the final cumulative mass of core and expanded atmosphere at the end of the first growth phase (see \citealt{Turrini2021} for further discussion), and $e$ is the Euler number.}

\textcolor{black}{The constant $\tau_{p}$ is the duration of the first growth phase and is set to about 1 Ma for both planets based on observational and theoretical constraints from circumstellar disks \citep{Manara2018,mulders2021,bernabo2022} and the Solar System \citep{scott2007,coradini2011,lichtenberg2022}. The individual values of $\tau_{p}$ are 1 Ma and 1.25 Ma for planets b and e, respectively, where the slower growth of planet e is introduced to delay the onset of its second growth and migration phase and allow planet b to get close to its final orbit before planet e reaches its peak rate of migration. This choice limits the chances of destabilizing close encounters between the two giant planets before they achieve their compact architecture.}

The second phase of mass growth accounts for the runaway gas accretion of the two giant planets, where their mass evolves as
%\begin{equation}
$ M_{p}(t)=M_{1}+\left( M_{2} - M_{1}\right)\left( 1-e^{-(t-\tau_{p})/\tau_{g}}\right)$
%\label{eqn-gasgrowth}
%\end{equation}
where M$_{2}$ is the final mass of the giant planets and $\tau_{g}$ is the e-folding time of the runaway gas accretion process. \textcolor{black}{Based on the discussion in Sect. \ref{sec:densities}, M$_{2}$ is set to 184 M$_\oplus$ for planet b when aiming to reproduce both the HMZe and LMZe scenarios from Table \ref{tab:scenarios}. In the case of planet e, M$_2$ is set to 135 M$_\oplus$ and 209 M$_\oplus$ when the simulations focus on the HMZe and LMZe scenarios, respectively.}%  when focusing on the LMZe scenario.}%when the focus is on the LMZe scenario.}% We will further discuss the impact of this choices for the planetary masses in the simulations in Sect. \ref{sec:conclusions}.} 
%and 180 M$_\oplus$, respectively, based on the estimated gas masses of the two giant planets discussed in Sect. \ref{sec:v1298tau}. 

The value of $\tau_{g}$ is set to 0.1 Ma based on the results of hydrodynamic simulations \citep{Lissauer2009,DAngelo2021}, meaning that the gas giants reach more than 99\% of their final mass in about 0.5 Ma from the onset of the runaway gas accretion. During the runaway gas accretion, %process,
giant planets form a gap in the disk gas whose width is modelled as $W_{gap} = C\cdot R_{H}$ \citep{Isella2016,marzari2018}, where the numerical proportionality factor $C=8$ is from \citet{Isella2016} and \citet{marzari2018} and $R_{H}$ is the planetary Hill's radius. The gas density $\Sigma_{gap}(r)$ inside the gap evolves over time with respect to the local unperturbed gas density $\Sigma(r)$ as $\Sigma_{gap}(r) = \Sigma(r)\cdot \exp{\left[-\left(t-\tau_{p}\right)/\tau_{g}\right]}$ \citep{Turrini2021}.\\

The physical radius of the growing giant planets ($R_{P}$) is a critical parameter governing the accretion efficiency of planetesimals. $R_{P}$ evolves together with the planetary mass across the two growth phases following the approach described by \citet{fortier2013}, which is based in turn on the hydrodynamic simulations of \citet{Lissauer2009}. During the first phase, the planetary core is growing its extended atmosphere and the physical radius evolves as:
%\begin{equation}
$ R_{P} = \frac{GM_P}{c_{s}^{2}/k_{1}+\left(GM_{P}\right)/\left(k_{2}R_{H}\right)}$
 %{\frac{c_{s}^{2}}{k_{1}}+\frac{GM_P}{k_{2}R_{H}}}$
% \label{eqn-inflatedradius}
%\end{equation}
where $G$ is the gravitational constant, M$_P$ is the instantaneous mass of the giant planet, $c_{s}$ is the sound speed in the protoplanetary disk at the orbital distance of the planet, $k_{1}=1$ and $k_{2}=1/4$ \citep{Lissauer2009}. 

When the giant planets enter their runaway gas accretion phase (i.e. for $t > \tau_{P}$), the gravitational infall of the gas causes the planetary radius to shrink as
$R_{P} = R_{E} - \Delta R \left(1-\exp^{-(t-\tau_{c})/\tau_{g}}\right)$
where $R_{E}$ is the planetary radius at the end of the extended atmosphere phase and $\Delta R = R_{E} - R_{T}$ is the decrease of the planetary radius during the gravitational collapse of the gas. We adopt as final values of the planetary radii $R_{T}$ the ones recently measured from TESS light-curves (see Table \ref{tab:v1298par}). Particles in the n--body simulations impact one of the giant planets when their relative distance from said planet is less than the planetary radius \citep[see][for further discussion]{Turrini2021}.

The migration of the giant planets over two growth phases is modeled after the migration tracks from \citet{Mordasini2015} following the parametric approach by \citet{Turrini2021}. During the first growth phase the planets undergo a damped Type I migration regime with drift rate \citep{Turrini2021}
%\begin{equation}
$\Delta v_{1} = \frac{1}{2}\frac{\Delta a_{1}}{a_{p}}\frac{\Delta t}{\tau_{p}}v_{p}$
%\label{eqn-typeImigration}
%\end{equation}
where $\Delta t$ is the timestep of the N-body simulation, $\Delta a_{1}$ is the radial displacement during the first growth phase, and $v_{p}$ and $a_{p}$ are the instantaneous planetary orbital velocity and semi-major axis, respectively. During the second growth phase, encompassing the transition to full Type I regime first and Type II regime later, the drift rate becomes \citep{hahn2005,Turrini2021} 
%\begin{equation}
$\Delta v_{2} = \frac{1}{2}\frac{\Delta a_{2}}{a_{p}}\frac{\Delta t}{\tau_{g}}\exp^{-\left(t-\tau_{c}\right)/\tau_{g}} v_{p}$
%\label{eqn-typeIImigration}
%\end{equation}
where $\Delta a_{2}$ is the radial displacement during this second phase (see Sect. \ref{sec:results-formation} for the values adopted in the simulations).

We set the spatial density of planetesimals in the N-body simulations to 1000 particles/au, with the inner edge of the planetesimal disk at 1 au and the outer edge at $r_c$ (see \citealt{Turrini2021} for the discussion of the choice of the disk inner edge). To compute the mass of heavy elements accreted by V1298\,Tau during their growth and migration, we treated each impacting particle in the N-body simulations as a swarm of real planetesimals. The cumulative mass of each swarm is computed integrating the disk gas density profile over a ring wide 0.1 au centered on the initial orbit of the impacting particle, and multiplying the resulting gas mass by the local solid-to-gas ratio. 

The solid-to-gas ratio is a function of the disk metallicity and local disk midplane temperature,  and is described by a simplified radial profile based on the realistic ones from \citet{Turrini2021} and \citet{pacetti2022}. The disk metallicity is set to 1.4\% \citep{asplund2009} based on V1298\,Tau's solar metallicity \citep{suarezmascarenoetal2021natast6_232}. The solid-to-gas ratio is 0.5 times the disk metallicity for planetesimals formed at temperatures comprised between 1200 K and 140 K, i.e. between the condensation of silicates and that of water. The solid-to-gas ratio grows to 0.75 times the disk metallicity for planetesimals formed at temperatures comprised between 140 K and 30 K, i.e. between the snowlines of water and carbon monoxide, and reaches 0.9 times the disk metallicity for planetesimals formed at temperatures below 30 K, i.e. beyond the carbon monoxide snowline.

\subsection{Resonant capture and resonance break simulations}\label{sec:methods-dynamics}
%\subsection{MEGNO and the resonant capture n--body simulations}
%{\color{green} 
%To evaluate the long-term stability of different planetary configurations, we exploit the chaos indicator MEGNO, the Mean Exponential Growth factor of Nearby Orbits \citep{MEGNO1,MEGNO2}. MEGNO is closely related to the maximum Lyapunov exponent and is based on solving the variational equations of the dynamical system to estimate the relative divergence of orbits. In the case of regular or quasi–periodic motion the MEGNO indicator is very close to 2, while for chaotic motion it increases with time. \textcolor{black}{The study by \cite{gajdos2023} confirms how the information supplied by MEGNO fully agrees with that arising from the Lyapunov time.}
%In the case of regular or quasi–periodic motion the MEGNO indicator is very close to 2, while for chaotic motion it increases with time. \textcolor{black}{The study by \cite{gajdos2023} shows how the information supplied by MEGNO fully agrees with that arising from the Lyapunov time. The study also shows how adopting a threshold MEGNO value of 3 between stable and chaotic systems is enough to avoid incorrect classifications, as the MEGNO value of chaotic systems is significantly larger than 2.}

To study the resonant capture process responsible for the present architecture of V1298\,Tau, we perform N-body integrations of V1298\,Tau's planets on converging orbits simulating their migration due to the interaction with the disk. To perform the n--body simulations we modified the RADAU algorithm \citep{Everhart85} to include different damping terms in semi--major axis, eccentricity, perihelion precession and velocity of the planet. 

\textcolor{black}{The parameters of these damping terms are time-dependent and can be tuned to simulate the disk dissipation timescales: to this end we adopt damping terms that decrease exponentially on a timescale of 0.5 Ma. The four planets are assumed fully formed (i.e. the resonant capture simulations take place after the conclusion of the formation simulations) and the disk close to its final dispersion. The masses adopted for the planets b and e are the nominal ones of Table \ref{tab:v1298par}. For the two inner planets, V1298\,Tau\,c and d, only upper limits on their mass values m$_c$<0.24 M$_J$ and m$_d$<0.31 M$_J$ are available \citep{suarezmascarenoetal2021natast6_232}.}

\textcolor{black}{We consider two mass configurations for these planets: in the first the planets are very light with a density of $\rho = 0.65$ $g/cm^3$, giving mass values of m$_c$=0.045 $M_{J}$ and m$_d$=0.077 $M_{J}$. This choice is based on the assumption that the planets are significantly puffed up due to their young age. We also consider the case with density $\rho = 1.3$ $g/cm^3$ for both planets. This case results in mass values $m_c$=0.09 $M_{J}$ and $m_d$=0.15 $M_{J}$, which are still within the observational constraints set by \cite{suarezmascarenoetal2021natast6_232} as well as the upper limits set by \cite{Tejada2022}.}

\textcolor{black}{We use the resonant configurations obtained with these simulations to select the starting conditions to study the break up of the resonant chain by planet-planet and planetesimal-planet scattering. The simulations of the interactions between massive planetesimal belts and V1298\,Tau's planets are performed with Mercury-Ar$\chi$es using a version of its hybrid symplectic algorithm ported to GPU computing through OpenACC.}
%}

\section{Results}\label{sec:results}

\subsection{Constraining the formation tracks of V1298\,Tau\,b and e}\label{sec:results-formation}

\textcolor{black}{The campaign of N--body simulations explores the characteristics of the circumstellar disk and the migration tracks that can give rise to the HMZe and LZMe scenarios of Table \ref{tab:scenarios}. Due to the uncertainty on the planetary masses and densities of V1298\,Tau\,b and e, %and the unknown contributions of their cores to their total metallicity, 
the goal of the simulations is to gather indications of how divergent the formation histories of the two giant planets were rather than to pinpoint their exact migration tracks. When comparing the planetesimals accreted in the simulations to the amounts of heavy elements reported in Table \ref{tab:scenarios}, we consider the two values to match when the differences are limited to %of the order of 
10-20\% (i.e. smaller than the uncertainty on the masses).}

%on two assumptions based on the results of \cite{Shibata2020}, \cite{Turrini2021} and \cite{pacetti2022}, and on the estimated amounts of heavy elements in V1298\,Tau\,e and b from Sect. \ref{sec:v1298tau}. First, the circumstellar disk surrounding V1298\,Tau should have originally contained at least $\sim$250 M$_\oplus$ of solids to be able to form the cores of the two giant planets and supply them with enough heavy elements during their migration (for simplicity we are not including here the solid material that formed V1298\,Tau\,c and d). Second, the initial orbital regions of the two giant planets should be far apart, with the initial orbit of V1298\,Tau\,e being a few times farther out than that of V1298\,Tau\,b, to allow the giant planets to encounter and accrete enough mass of planetesimals.

\textcolor{black}{We consider three host circumstellar disks (see Table \ref{tab:formation-scenarios}). The first disk has mass of 0.06 M$_\odot$, comparable to that of a Minimum Mass Solar Nebula-like disk \citep{hayashi1981} with the radial extension reported in Sect. \ref{sec:model-formation} (simulations 1-4 in Table \ref{tab:formation-scenarios}). The second and third disks are twice (0.12 M$_\odot$, simulations 5-12 in Table \ref{tab:formation-scenarios}) and three times (0.18 M$_\odot$, simulations 13-18 in Table \ref{tab:formation-scenarios}) more massive, respectively. The increasing masses of these disks impact the planetesimal accretion efficiencies of the planets both by providing more solid material and by changing the balance between gas drag and planetary perturbations. As a consequence, the results do not simply scale linearly with the amount of solid material available.}

%The first scenario takes place in a disk with mass comparable to that of a Solar Nebula with the same extension \citep{hayashi1981}. While this disk does not have enough mass in solids to produce planets with the desired enrichment in heavy elements, this scenario is quite informative on the effects of the concurrent planetesimal accretion of the two planets. If V1298\,Tau\,b is too efficient in accreting planetesimals, the feeding zone of V1298\,Tau\,e becomes planetesimal-depleted and the two planets end up with similar enrichments notwithstanding their different formation regions.

\begin{table*}[]
    \renewcommand{\arraystretch}{1.2}
    %The height of each row is set to 1.5 relative to its default height.
    \centering
    \begin{tabular}{|c | c c c | c c c c | c c c c|}
    \hline
    Id. &\multicolumn{3}{c|}{Disk} & \multicolumn{4}{|c|}{V1298\,Tau\,b} & \multicolumn{4}{|c|}{V1298\,Tau\,e}\\
    \hline
     & M$_{\text{disk}}$   & $\Sigma_{0}$ & M$_{dust}$& Initial & $\Delta a_1$ & $\Delta a_2$ & Accreted & Initial & $\Delta a_1$ & $\Delta a_2$ & Accreted \\
     & (M$_\odot$) & (g/cm$^3$) & (M$_\oplus$) & Seed (au) & (au) & (au) & Mass (M$_\oplus$) & Seed (au) & (au) & (au) & Mass (M$_\oplus$)\\
    \hline
    %18
    1 & 0.06  & 42 & 205 & 11 & -6.4 & -4.3 & 22 & 50 & -29.6 & -19.8 & 54 \\
    %1
    2 & 0.06  & 42 & 205 & 10 & -5.8 & -3.9 & 19 & 30 & -17.6 & -11.8 & 46 \\
    %8
    \rowcolor[HTML]{e1f5fe}
    3 & 0.06  & 42 & 205 & 10 & -5.8 & -3.9 & 18 & 20 & -11.6 & -7.8 & 26 \\
    %9
    \rowcolor[HTML]{e1f5fe}%{fffde7}
    4 & 0.06  & 42 & 205 & 11 & -6.4 & -4.3 & 21 & 22 & -12.8 & -8.6 & 28 \\
    %
    % DUST PRODUCTION
    % 5.21694E+01   3.06411E-01   7.86223E+01   4.61780E-01
    \arrayrulecolor{gray}
    \hline
    \arrayrulecolor{black}
    %2
    5 & 0.12  & 80 & 390 & 6 & -3.4 & -2.3 & 15 & 50 & -29.6 & -19.8 & 139 \\
    %3
    6 & 0.12  & 80 & 390 & 6 & -3.4 & -2.3 & 13 & 15 & -8.6 & -5.8 & 52 \\
    %5
    7 & 0.12  & 80 & 390 & 9 & -5.2 & -3.5 & 30 & 50 & -29.6 & -19.8 & 132 \\
    %6
    8 & 0.12  & 80 & 390 & 9 & -5.2 & -3.5 & 36 & 12.5 & -7.1 & -4.8 & 11 \\
    %11
    \rowcolor[HTML]{f9fbe7}
    9 & 0.12  & 80 & 390 & 7.5 & -4.3 & -2.9 & 20 & 50 & -29.6 & -19.8 & 129 \\
    %
    %DUST PRODUCTION
    % 1.56516E+02   4.63013E-01   2.74790E+02   8.12895E-01
    %12
    10 & 0.12  & 80 & 390 & 7.5 & -4.3 & -2.9 & 19 & 15 & -8.6 & -5.8 & 41 \\
    %13
    \rowcolor[HTML]{e1f5fe}
    11 & 0.12  & 80 & 390 & 7.5 & -4.3 & -2.9 & 20 & 12.5 & -7.1 & -4.8 & 25 \\    
    %17
    \rowcolor[HTML]{e1f5fe}%{fffde7} 
    12 & 0.12  & 80 & 390 & 7.5 & -4.3 & -2.9 & 19 & 13 & -7.4 & -5.0 & 28 \\
    \arrayrulecolor{gray}
    \hline
    \arrayrulecolor{black}
    %4
    13 & 0.18  & 120 & 585 & 5 & -2.8 & -1.9 & 7 & 50 & -29.6 & -19.8 & 250 \\
    %7
    14 & 0.18  & 120 & 585 & 10 & -5.8 & -3.9 & 48 & 50 & -29.6 & -19.8 & 233 \\
    %10
    15 & 0.18  & 120 & 585 & 7.5 & -4.3 & -2.9 & 26 & 50 & -29.6 & -19.8 & 234 \\    
    %14
    \rowcolor[HTML]{fff3e0}
    16 & 0.18  & 120 & 585 & 7 & -4 & -2.7 & 23 & 50 & -29.6 & -19.8 & 245 \\ 
    %15
    17 & 0.18  & 120 & 585 & 7 & -4 & -2.7 & 23 & 12.5 & -7.1 & -4.8 & 42 \\ 
    %16
    \rowcolor[HTML]{e1f5fe}%{fffde7}
    18 & 0.18  & 120 & 585 & 7 & -4 & -2.7 & 23 & 11 & -6.2 & -4.2 & 28 \\
    \hline
    \end{tabular}
    \caption{Parameters of the circumstellar disks and the migration scenarios considered in the 18 formation simulations for V1298\,Tau's planets b and e, and the resulting accretion of heavy elements of each scenario. \textcolor{black}{The rows highlighted in blue identify the simulations consistent with the LMZe scenario in Table \ref{tab:scenarios}, the row highlighted in red the simulation consistent with the HMZe scenario, while the green row identifies the simulation that can be consistent with the HMZe scenario assuming a second process assists in increasing the metallicity of planet e.}}
    \label{tab:formation-scenarios}
\end{table*}

\begin{figure*}
    \centering
    \includegraphics[width=\textwidth]{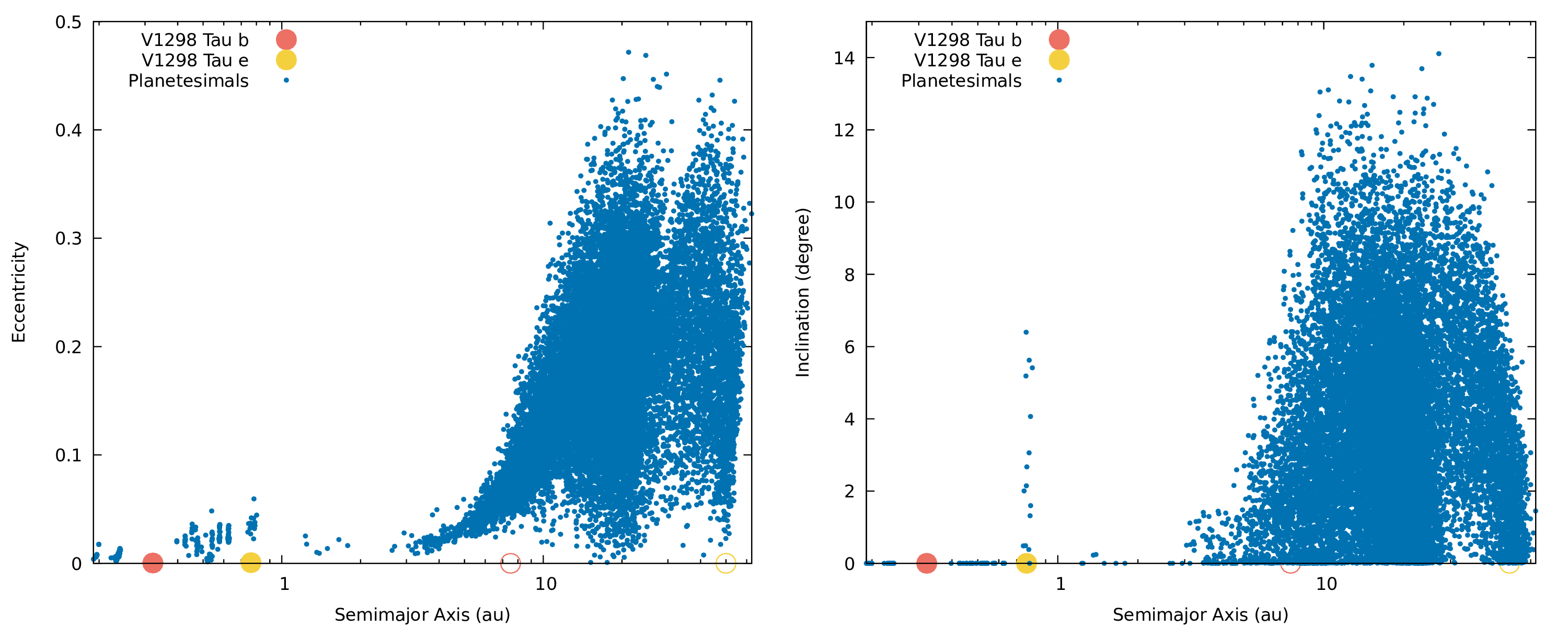}%{mosaic-fei.png}
    \caption{Dynamical excitation of the planetesimal disk (\textit{left}: eccentricity, \textit{right}: inclination) after the formation and migration of V1298\,Tau\,b and e in scenario 9 from Table \ref{tab:formation-scenarios}. Empty circles mark the initial position of the giant planets. The simulation stopped with the giant planets about 3x more distant from V1298\,Tau than their current orbits to prevent dynamical instabilities that would result in a system incompatible with V1298\,Tau's compact architecture.}
    \label{fig:formation-snapshot}
\end{figure*}

\textcolor{black}{Table \ref{tab:formation-scenarios} shows the combination of disk parameters, formation regions and migration scenarios we explored. As illustrated by Table \ref{tab:formation-scenarios}, consistently with the results of \cite{Shibata2020} the amounts of heavy elements accreted by the giant planets can be enhanced or reduced by proportionally increasing or decreasing their orbital displacements, respectively (see Sect. \ref{sec:conclusions} for further discussion). The partition of the displacements between the two migration phases is chosen to maximise the planetesimal accretion efficiency of the planets  \citep{Shibata2020,Turrini2021}.} 

\textcolor{black}{The total mass of solids contained in the first disk (0.06 M$_\odot$) is not enough to support the HMZe scenario, so in simulations 1-4 we focus exclusively on the LMZe scenario. The metallicity values of the two giant planets in the LMZe scenario are obtained when V1298\,Tau\,b begins its formation around 10 au and V1298\,Tau\,e about twice as distant as showcased by simulations 3 and 4. This picture is qualitatively preserved in the two more massive disks. The LMZe scenario is reproduced for V1298\,Tau\,b starting its formation slightly closer to the star (about 7 au in both disks) and V1298\,Tau\,e being initially located 1.5-2 times farther away (see simulations 12 and 18 in Table \ref{tab:formation-scenarios}).} 

\textcolor{black}{The HMZe scenario can be reproduced only by the most massive disk considered in our simulations (0.18 M$_\odot$). Even in this case, the high metallicity value of V1298\,Tau\,e requires the planet to start its formation at the outer edge of the disk in order to encounter and accrete enough planetesimals (see simulations 16). In the intermediate disk (0.12 M$_\odot$), the same extreme migration track results in about half the metallicity required for V1298\,Tau\,e in the HMZe scenario (see simulation 9 in Table \ref{tab:formation-scenarios}). An additional process, e.g. the accretion of disk gas enriched in heavy elements \citep[e.g.][]{booth2019,cridland2019,schneider2021}, needs to be invoked together with planetesimal accretion to explain the missing metallicity.}

\textcolor{black}{The larger enrichment required in case of the lower density values of the inflated gaseous envelopes expected for the two hot and young giant planets (87 and 108 M$_\oplus$ for V1298\,Tau\,b and e in the LMZe scenario, see Sect. \ref{sec:densities}) point to both planets having undergone extensive migration in a disk whose mass was at least 10\% that of the host star (i.e. similar or more massive than our intermediate disk). The results reported in Table \ref{tab:formation-scenarios} suggest that, in this scenario, V1298\,Tau\,b should have started its formation about midway through its host disk and V1298\,Tau\,e about twice as far (i.e. closer to the outer edge of the disk). Lower planetary mass values than the nominal ones reported in Table \ref{tab:v1298par} or the presence of massive cores (which we ignored in our computations) would instead lower the required planetesimal accretion and shift inward the initial formation regions of the planets.}

\textcolor{black}{As the previous results and discussion showcase, the planetary density and metallicity values allow the discussion of the formation histories of the two giant planets only from a comparative point of view. Specifically, the picture emerging from our campaign of simulations consistently indicates that V1298\,Tau\,e should have formed further out than V1298\,Tau\,b, likely about twice as far from their host star. The uncertainty on their masses and our ignorance of the characteristics of their now long-gone native disk, however, prevent us from precisely pinpointing the formation regions of the two giant planets. We will further discuss this issue in Sect. \ref{sec:discussion-formation} and highlight how the future atmospheric characterization of the two planets will allow us to overcome it in Sect. \ref{sec:discussion-atmospheres}.}

%Before proceeding, it is worth pointing out that the longer orbital period of V1298\,Tau\,e suggested by the observations of TESS \citep{feinsteinetal2022apjl925_2} may require the revision of the planetary mass value estimated through RV measurements by \cite{suarezmascarenoetal2021natast6_232}. As illustrated by the scenarios reported in Table \ref{tab:formation-scenarios}, a lower density value of planet e would allow for a less massive circumstellar disk or could indicate that the giant planet formed closer to its host star and underwent less extensive migration. However, such a downward revision of the mass and density values of V1298\,Tau\,e would not affect qualitatively the formation and dynamical scenarios discussed here and in the following sections. 

\subsection{Impact of the formation tracks of V1298\,Tau\,b and e on the subsequent evolution of the system}\label{sec:results-dust}

As shown in Fig. \ref{fig:formation-snapshot}, the migration histories of V1298\,Tau\,b and e shape the characteristics of the surviving planetesimal disk in two ways.  First, the migration of the two giant planets leaves in its wake surviving planetesimal disks that are characterized by two different dynamical regions (see Fig. \ref{fig:formation-snapshot}). The innermost region, roughly extending from the current orbits of the giant planets to the initial formation region of V1298\,Tau\,b (see Fig. \ref{fig:formation-snapshot}), is depopulated of planetesimals. Furthermore, due to the higher gas density in the inner disk regions, the dynamical excitation of the surviving planetesimals is efficiently damped by gas drag. The outermost region, roughly extending between the initial formation regions of V1298\,Tau\,b and e (see Fig. \ref{fig:formation-snapshot}), is populated by dynamically excited planetesimals.

Second, the \textcolor{black}{larger} migration of V1298\,Tau\,e favours its acquisition of a marked population of Trojan satellites (see the planetesimals ``piled up'' on V1298\,Tau\,e in Fig. \ref{fig:formation-snapshot}), planetesimals co-orbiting with the giant planet in stable orbits around its L4 and L5 Lagrangian points with the star, similarly to what is argued in the case of the Solar System for Jupiter \citep{pirani2019}. The formation history of V1298\,Tau\,b appears to hinder its efficiency in capturing Trojan satellites, although this could be an artefact of the specific migration tracks we adopt. %The simulations we performed stop when the giant planets arrived about three times farther away from the star than their current orbits to prevent the triggering of dynamical instabilities in the final phases of migration, since the damping effect of the disk on giant planets is not included in the simulations. 
Dedicated studies of the convergent migration and resonant trapping of V1298\,Tau\,b and e are required to assess the dynamical lifetime of possible Trojan bodies.

The population of dynamically excited planetesimals created by the forming giant planets can affect the later evolution of the planetary system in two ways. On the one hand, the resulting higher impact frequency among planetesimals promotes the growth of existing oligarchs and the formation of additional massive planetary bodies\textcolor{black}{, whose mutual interactions can inject them on eccentric orbits} outside the present orbits of the four known planets. Later encounters between such excited outer planets and the four resonant inner planets could result in the break-up of the resonant chain and the onset of dynamical instability. On the other hand, the high-velocity impacts between the excited planetesimals cause large-scale production of collisional dust, replenishing the dust population of the circumstellar disk as shown by the results of \cite{Turrini2019} and \cite{bernabo2022}.

We used the \textsc{Debris} code \citep{Turrini2019,bernabo2022} to estimate the possible amount of collisional dust produced in \textcolor{black}{simulations 4 and 9  (see Table \ref{tab:formation-scenarios} and Fig. \ref{fig:formation-snapshot} for the excited planetesimal disk at the end of simulation 9) by the migration of the V1298\,Tau b and e.} We assume the planetesimal population to be characterized by the size-frequency distribution and mechanical strength typical of primordial planetesimals formed by streaming instability \citep{Krivov2018,Turrini2019}. We refer interested readers to \cite{Turrini2019} and \cite{bernabo2022} for details on the collisional and dust production algorithms implemented in the \textsc{Debris} code. 

The collisional cascade among the surviving excited planetesimals is capable of converting back into dust \textcolor{black}{between 50 M$_\oplus$ (simulation 4) and 160 M$_\oplus$ (simulation 9). Such an amount of collisional dust, while drifting inward toward the star, is} large enough to impact the formation history of V1298\,Tau's system. First, it promotes pebble accretion on surviving planetary embryos not accreted by the migrating planets, resulting in the formation of additional massive planets. We will discuss the role of such planets in shaping the current architecture of V1298\,Tau in Sect. \ref{sec:planet-planet_scattering}. Second, it promotes new phases of streaming instability and can form massive planetesimal belts outside the current orbits of V1298\,Tau\,e. The gravitational interactions between such belts and V1298\,Tau's four resonant planets can also potentially break the resonance chain. We will investigate this scenario in Sect. \ref{sec:planetesimal_perturbations}.

\subsection{Primordial capture in 4--body resonance as V1298\,Tau's original architecture}\label{sec:capture}

\textcolor{black}{As discussed in Sect. \ref{sec:namd}, the current architecture of V1298\,Tau's planets is not characterized by the resonant chain expected for such compact systems and shows the signs of having been sculpted by dynamical instabilities. To understand the origin of these characteristics, we first explore the possible primordial architectures of the system. } There are two possible Laplace resonance chains which are dynamically close to the nominal solution and stable: they are the 3:2, 2:1, 3:2 chain and the 3:2, 2:1, 2:1 chain.  

A possible scenario for the formation of either of the two resonant chains is that the planets migrated into resonance while the circumstellar disk was still present\textcolor{black}{, consistently with the migration scenarios simulated in Sect. \ref{sec:results-formation} to explain the density values of V1298\,Tau\,b and e}. \textcolor{black}{The capture in both these Laplace resonances is simulated with the parallel version of the N--body integration code RADAU15 \citep{Everhart85} described in Sect. \ref{sec:methods-dynamics} using exponential damping terms in semi--major axis and eccentricity with e-folding times of 0.5 Ma.} 

\textcolor{black}{The masses adopted for the planets b and e are the nominal ones of Table \ref{tab:v1298par} while for planets c and d we consider two configurations (see Sect. \ref{sec:methods-dynamics}): a first one where the planets have density $\rho = 0.65$ $g/cm^3$ and masses m$_c$=0.045 $M_{J}$ and m$_d$=0.077 $M_{J}$, and a second one where they have density $\rho = 1.3$ $g/cm^3$ and masses $m_c$=0.09 $M_{J}$ and $m_d$=0.15 $M_{J}$. We find no significant differences between the two configurations.}

The way in which the real capture in resonance may have occurred depends on a large number of parameters, among which: the initial orbits of the planets, the disk gas density profile and the timescale of its dissipation, the local effects of the planets on the gas, the viscosity of the disk and its temperature profile. As it is impossible to perform the complete exploration of all these parameters, we first find a case in which the trapping occurs to confirm that it is indeed a realistic scenario. 

Then, starting from this single case, we explore the phase space nearby this resonant solution to find all possible resonant configurations. This approach allows to map the range of the planetary orbital elements in the specific Laplace resonant configuration. A random search without the knowledge of the orbital elements of at least one resonant configuration would be impossible. Depending on the above mentioned parameters influencing the planet migration rate, different resonant configurations, among those we have found, will be achieved during the evolution of the system.

\textcolor{black}{The detailed exploration of the resonant phase space is performed by randomly sampling all planetary orbital elements in the proximity of the resonant solution. During the numerical integration of the planetary orbits, %with a pure N--body code, 
the critical angles of the Laplace resonance are automatically checked and the non-librating, non-apsidal corotation cases are rejected. This last condition is dictated by the results of \cite{beauge2006} suggesting that, as long as the migration is sufficiently slow to be approximated as an adiabatic process, all captured planets must be in apsidal corotations. The range we find in eccentricity shows the possible final orbital configurations of the four planets at the end of migration and resonance capture for potentially different disk and planet initial parameters.}

In  Fig.~\ref{fig:migra32} we illustrate the trapping in the 3:2, 2:1, 3:2 resonant chain where the planets are started  at the beginning of the simulation close to the resonance and they get captured during the subsequent inward migration. The eccentricity of each planet at the resonance trapping is pumped up to a value which is kept constant till the end of migration. These high values are compatible with observations.  The critical arguments of the three individual resonances are librating around different values. In Fig.~\ref{fig:migra21} we show the second scenario of resonance trapping where the planets are captured in a  3:2, 2:1, 2:1 resonant chain. Even in this case, the eccentricity is pumped up and the resonant angles are  all librating.  It must be noted that these simulations can be scaled in semi--major axis to get as close as possible to the observed  semi--major axes of the planets. 

{\color{black} In the subsequent detailed explorations of the phase space we numerically integrate  millions of initial planetary configurations to find the extension of the phase space of the Laplace resonant configurations. 
In Fig.~\ref{fig:reso_wide}
we show the ranges of the initial values of semi-major axis (very tiny) and  eccentricity with all four planets locked in a Laplace resonance in both cases (3:2, 2:1 and 3:2 on the left and
3:2, 2:1 and 2:1 on the right). Within these ranges there are resonant solutions which are compatible in terms of eccentricity with the observed nominal system, suggesting that the primordial system was indeed in a Laplace resonance and it has later %just 
escaped from the resonant configuration. % in recent times.} 
By comparing the two different configurations, 3:2, 2:1 and 3:2 vs. 3:2, 2:1 and 2:1, it is also interesting to note that the last configuration leads to higher eccentricities for the third planet. 

\textcolor{black}{It is worth pointing out that the resonant solutions plotted in Fig.~\ref{fig:reso_wide} are obtained by requiring that both resonant arguments librate, a conditions naturally leading to apsidal libration. This condition is in agreement with the work of \cite{beauge2006} who  suggest that, as long as the orbital migration is sufficiently slow to be approximated by an adiabatic process (a condition compatible with the slow migration of giant planets toward the end of their formation, see e.g. \citealt{pirani2019,tanaka2020}), all captured planets should be in apsidal corotations.}

\begin{figure*}
\centering
\includegraphics[width=\textwidth]{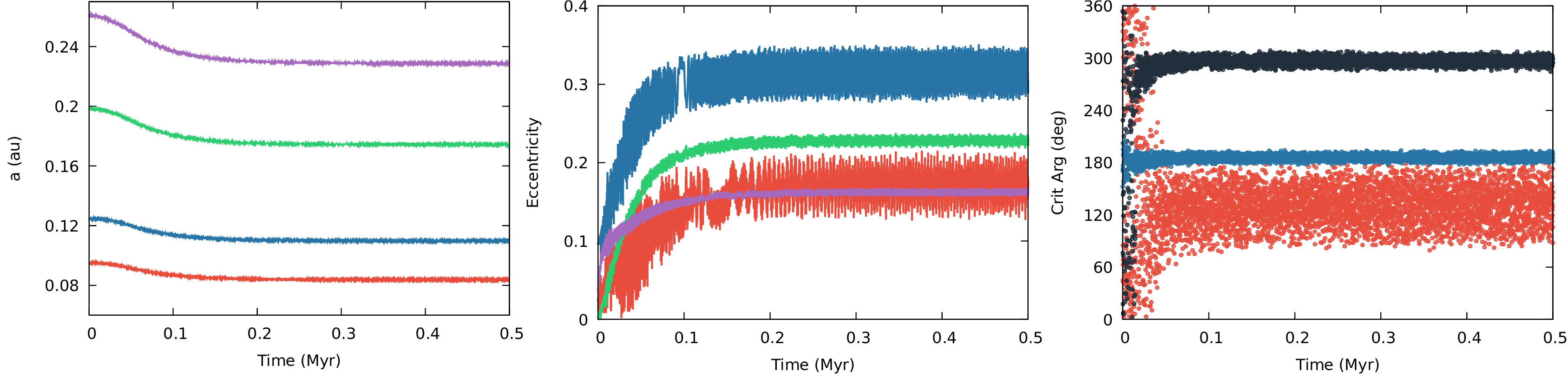}
\caption{Migration of the four planets in resonant configuration with the planets in the 3:2, 2:1 and 3:2 resonances. The left panel shows the semi--major axis evolution, the middle panel the eccentricity and the right panel the resonant critical angles during the migration and subsequent resonance capture. The gas is slowly dissipating across the simulation.}\label{fig:migra32}
\end{figure*}

\begin{figure*}
\centering
\includegraphics[width=\textwidth]{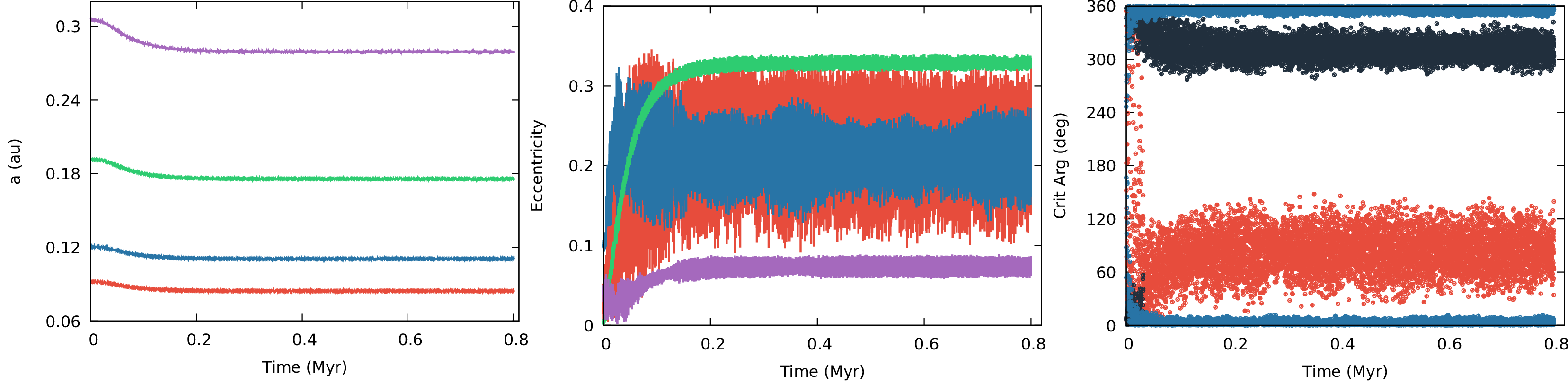}
\caption{Same as Fig.~\ref{fig:migra32} but for the resonance sequence 3:2, 2:1, 2:1.}
\label{fig:migra21}
\end{figure*}

%Portatile: /home/francesco/V1298/MIGRA\_TURRINI/FASTER\_OK

\begin{figure*}
\centering
\includegraphics[width=0.66\textwidth]{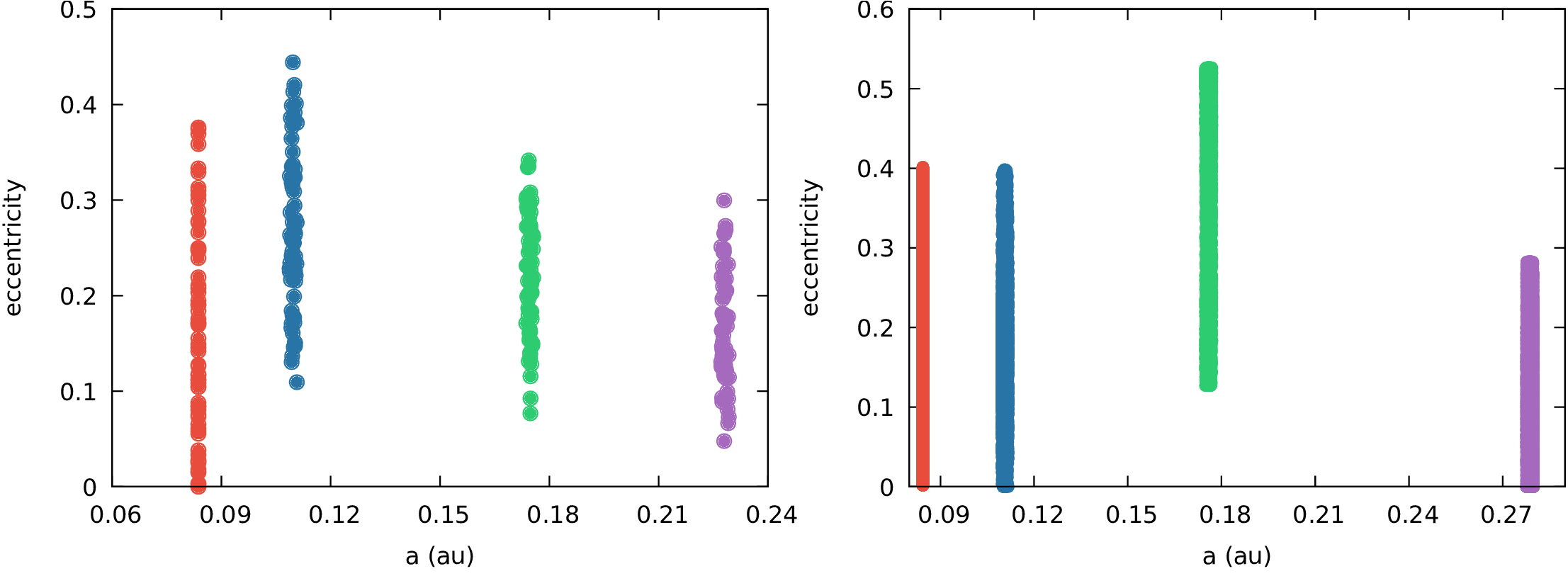}
\caption{\label{fig:reso_wide} Random resonant solutions in the semi--major axis and eccentricity planet for the two different chains (3:2, 2:1 and 3:2 on the left and 3:2, 2:1 and 2:1 on the right). The solutions are scalable in semi--major axes and the three inner planet semi--major axes can be shifted until they coincide with the observed ones.}
\end{figure*}

%voyager: /home/marzari/V1298/RESONANCE\_WIDTH
Even if the resonant lock provides a solution for %the stability problem of 
the origin of V1298\,Tau's compact architecture, at present the system is not in resonant chain \textcolor{black}{as confirmed by the extensive parameter exploration performed by \cite{Tejada2022}}. The ratios between the planetary semi--major axes, according to the current nominal solutions from Table \ref{tab:v1298par}, are \textcolor{black}{1.312, 1.559, 1.535 while to be in a 4--body resonance they should be 1.310, 1.587,  1.310 or 1.310, 1.587, 1.587.} %1.311534,1.558477, 1.535195 while to be in a 4--body resonance they should be 1.31037, 1.5874,  1.31037 or 1.31037, 1.5874, 1.5874. 
%\textcolor{black}{The current non-resonant architecture of the system is confirmed by the extensive parameter exploration performed by \cite{Tejada2022}.} %It is possible that 

\textcolor{black}{This mismatch can be explained if} the system evolved into either of the two resonant chains in the initial stages of its formation history but, during its later evolution, the resonance lock was broken by some additional dynamical mechanism. %leaving the planets in the unstable state presently observed. 
According to the capture simulations with a dissipating disk  (Fig.~\ref{fig:migra32} and Fig.~\ref{fig:migra21}), it is difficult for the gas dispersal alone to break the resonance lock. Dynamically destabilizing mechanisms must be invoked, specifically the planet-planet scattering and planetesimals scattering introduced in Sect. \ref{sec:results-formation} and explored in detail in the next sections. 

%However, we must take into account that there are error bars in the values of the semi--major axes.
%Generating random systems within the errors (Gaussian distribution within 3 sigma) and comparing the systems to resonance widths, 
%we find that only about 140 over $1 \cross 10^6$ systems are
%located within the resonance widhts. 
%However, the resonance widths can be different because of the Laplace resonance between the 4 planets. 

%\begin{figure}
%\centering
%\includegraphics[width=0.3\textwidth,angle=-90]{noreso.pdf}
%\caption{\label{fig:nominal2} In magenta the systems which are within the resonance widths of the resonant chain.}
%\end{figure}

%portatile:  /V1298/RESO\_NORESO
\subsection{V1298\,Tau's present architecture: resonance break by planet--planet scattering}\label{sec:planet-planet_scattering}

The first possible scenario to break the resonant lock requires that additional dynamically excited planets populated, and maybe still populate, the system on orbits external to that of planet e. These excited planets would be difficult to detect through transits due to their longer orbital periods. % and inclined orbits. 
As discussed in Sect. \ref{sec:results-dust}, the convergent migration shaping the formation path of V1298\,Tau could act to promote the formation of such planets. \textcolor{black}{These additional planets would have formed at later times in the wake of the passage of V1298\,Tau\,b and e (see Sect. \ref{sec:results-dust}) and would not migrate as close to the star as them} either because of the dissipation of the gaseous disk or because they began to interact gravitationally with each other \citep{marzari2010}. The dissipation of the disk itself can cause the onset of the dynamical instability among the resulting population of planets. 

During their mutual gravitational interactions, one (or more) of such outer planet may have ended up in a highly eccentric orbit (\citealt{weimar1996,rasioford1996,lin1997} and \citealt{davies2014} for a review \textcolor{black}{ of the dynamical process and \citealt{limbach2015,zinzi2017,Turrini2020,turrini2022} for its signatures in observed multi-planet systems}). This additional planet on an eccentric orbit %, outcome of planet--planet scattering with the outer planets,  
may get close enough to the four presently discovered planets while they were locked in a Laplace resonance and destabilize them. Encounters between this additional planet and planet e would break the resonance chain leading the system of inner planets to instability on short timescales.  The young age of the system suggests that the event triggering the instability of the outer planets, from which the fifth planet originated, could have been the dissipation of the gaseous disk, unless the planet formed directly on an eccentric orbit from the planetesimal disk excited by the migration of planets b and e. 

In Fig.~\ref{jumping} we show two examples of such planet-planet scattering scenario: in the first case (top panel) a 200 M$_{\oplus}$ planet is added to the system on an orbit with $a=8 $ au, $e=0.95$ and $i = 10^o$. The inner planets remain locked in the 4--body resonance until a series of close encounters with the fifth planet breaks the resonance lock, the critical arguments begin to circulate and the system is quickly destabilized. In the second case (bottom panel) a heavier planet with $M= 2 M_J$ is placed on an orbit with semi--major axis $a= 4 $ au, eccentricity $e=0.85$ and inclination $i = 2^\circ$ and a similar evolution is observed. 

These two examples are representative of the large sample of simulations we performed including outer planets with  mass ranging from $m$ = 150 M$_{\oplus}$ to $m$ = 2 M$_J$ and different initial eccentricities and semi--major axes, all leading to resonance break and instability.
The first case has a very high eccentricity which compensates the lower planet mass and the larger semi--major axis. It illustrates a scenario where the outer planetary system formed far away from the inner one and, becoming unstable, produced the perturbing planet. In the second case the perturbing planet is closer but with a smaller eccentricity and a larger mass, representing a scenario where the inner and outer planets where potentially born more packed. 

There is an infinite number of possible configurations of a perturbing planet on a crossing orbit with respect to the inner bodies leading to planet-planet scattering, chaotic evolution and instability. Due to the chaotic nature of the problem there is little reason for performing more extended explorations of the parameter space, particularly since our goal is solely to prove that this is a viable scenario. There is no way to identify the exact configuration that led to the present planetary system as even those closely reproducing the observed system would not be unique. The time--scale for the onset of instability depends on the initial conditions and can be tuned by changing the initial semi--major axis of the fifth planet to delay the instability. 

This behavior can explain why at present we observe V1298\,Tau's four planets close to a resonance chain but not locked in it. %, and doomed to soon undergo additional planet--planet scattering which will further change their orbital elements and fully destabilize them. 
and is supported by the high value of the NAMD of V1298\,Tau (see Sect. \ref{sec:v1298tau}), which is highly suggestive of a period of violent planetary encounters \citep{carleo2021,turrini2022}. Unfortunately, it is not possible to trace back with precision the initial conditions which led to the present unstable system even with a large sample of numerical simulations. The chaotic evolution due to planet-planet scattering can drive any putative system close to the observed one at different times during its evolution and many different initial conditions can bring the system close to the observed one. 

A possible alternative mechanism which may drive  a system initially formed in a resonant chain to instability is tidal eccentricity damping which may lead to divergence of the orbital semimajor axes of the resonant bodies \citep{bat2013,lit}. However, the eccentricity damping timescale of the innermost planet (planet c) is two orders of magnitude longer than the age of the system according to an estimate based on the tidal model of \cite{leconte2010} and a modified tidal quality factor of the planet of $10^5$. The mass loss of the innermost planets due to photoevaporation caused by their proximity to the star is an additional plausible candidate for triggering dynamical instabilities in a packed resonant chain \citep{gold2022}. Unfortunately, to be effective also this mechanism requires a timescale which is of the order of 100 Ma or more, significantly longer than the age of V1298 Tau (see Sect. \ref{sec:v1298tau} and Table \ref{tab:v1298par}).} \textcolor{black}{Finally, it is worth noting that the two inner planets are also very close to a 2:1 resonance and it is the third that is far from the resonant ratio with respect to the inner two.}

\begin{figure}[htbp]
\centering
\includegraphics[width=\columnwidth]{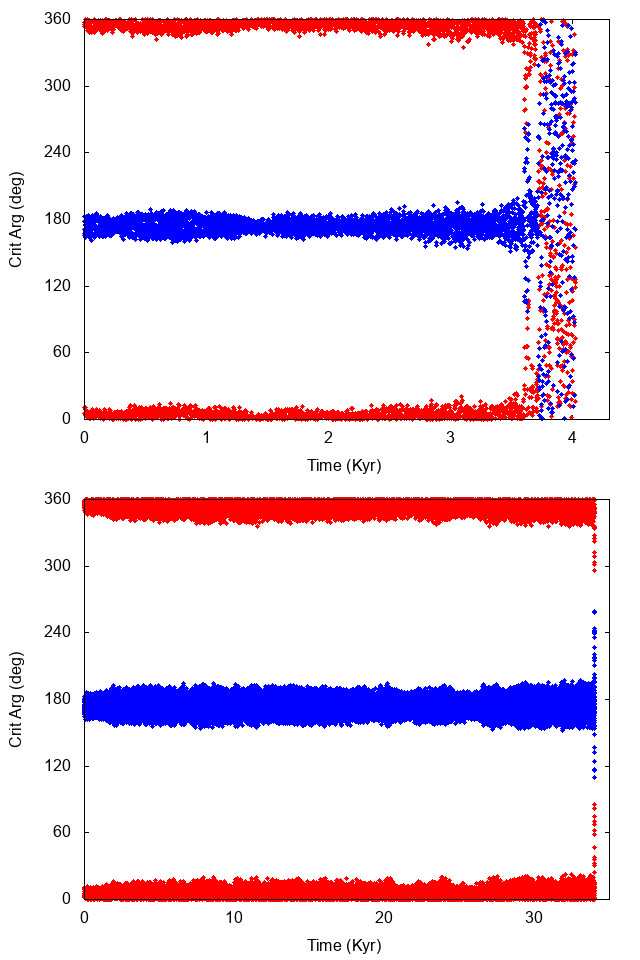}
\caption{\label{jumping} 
Evolution of the critical argument of the outer planet pair (b and e) initially locked in a 3:2 resonance. In the upper panel the additional planet has a mass  $m= 200 M_{\oplus}$, $a=8 $ au, $e=0.95$ and $i = 10^\circ$. In the bottom panel a more massive planet on a less eccentric orbits is adopted ($m$= 2 M$_J$, $a$= 4 au, $e$=0.85 and inclination $i = 2^o$). In both cases, after a period of chaotic evolution, the onset of repeated close encounters breaks the resonance lock and lead to fast instability. }
\end{figure}

\subsection{V1298\,Tau's present architecture: can planetesimal scattering break the resonance?}\label{sec:planetesimal_perturbations}

We investigate also the possibility that residual planetesimal scattering may be responsible for the resonant chain breaking and the current unstable configuration of the system. Remnant planetesimals may be present in the region where the planets migrate while in resonance. Furthermore, as  discussed in Sect. \ref{sec:results-dust}, the migrating giant planets leave in their wake  dynamically excited planetesimals whose high-velocity impacts will convert a significant fraction of their mass into second-generation dust and pebbles \citep{Turrini2012,Turrini2019,bernabo2022}. While drifting toward the star, this second-generation dust can trigger new phases of planetesimal formation and create new planetesimal belts closer to the resonant planets.

We consider two scenarios characterized by planetesimal belts of 10 and 50 M$_\oplus$, respectively, both extending between 0.3 au and 10 au. This is the same orbital region where gas drag efficiently damps the dynamical excitation of the planetesimals (see Fig. \ref{fig:formation-snapshot}): as a result, the planetesimals populating the belts start their evolution after the dispersal of the circumstellar disk on low-eccentricity, low-inclination orbits ($e\approx i<0.01$ where the inclination is expressed in radians). 

The 10 and 50 M$_\oplus$ planetesimals belts are simulated using 1000 massive particles whose initial masses are 0.01 and 0.05 M$_\oplus$, respectively. Collisions among planetesimals are treated as inelastic mergers and result in mass growth. The giant planets are placed in a stable resonant chain at the beginning of the simulations and the whole system is evolved for 10 Ma (i.e. the lower bound to V1298\,Tau's age in Table \ref{tab:v1298par}) with timestep of 0.35 days. The simulations are performed using the GPU-accelerated version of \texttt{Mercury-Ar$\chi$es} N-body code \citep{Turrini2019,Turrini2021} discussed in Sect. \ref{sec:methods-dynamics}.

\textcolor{black}{The simulations reveal that the presence of the 10 M$_\oplus$ planetesimal belt has negligible effects on the dynamical evolution of the four planets, which is the same as in the reference scenario with no planetesimal belt. The presence of the 50 M$_\oplus$ planetesimal belt produces limited alterations to the dynamical evolution of the planets, as the encounters with the more massive planetesimals slightly damp the eccentricity of V1298\,Tau\,e and reduce the amplitude of the eccentricity oscillations of V1298\,Tau\,b. Aside from these limited differences, the secular variations of the orbital elements of the four planets are within the ranges of the scenarios with the 10 M$_\oplus$ belt and no planetesimal belt. Overall, even the presence of the 50 M$_\oplus$ planetesimal belt does not appear capable of breaking the resonance chain over the lifetime of the system. Unless V1298\,Tau hosts a planetesimal belt more massive than those considered in this analysis, planetesimal scattering does not appear as a viable solution to break the resonance chain.}

\section{Observational clues on the presence of additional planets at wide separation}\label{sec:observations}

\textcolor{black}{The scenario of high enrichment in heavy elements of V1298\,Tau\,b and e we explore in Sect \ref{sec:results-formation} is} best reproduced if the two planets form and migrate across compact protoplanetary disks with radial extension <50-100 au. Wider disks imply lower spatial densities of the planetesimals as the disk mass is spread over wider areas since, for the same total disk mass, they would have a lower coefficient of the power law describing the mass density profile. This, in turn, implies that migrating giant planets encounter fewer planetesimals and accrete lower masses of heavy elements over the same radial displacement. 

Fitting the density values of planets b and e in such wide disks would require initial formation regions for the two planets similar to those of the giant planets revealed by ALMA surveys \citep[e.g.][]{Long2018,andrews2018}. However, it is currently unclear if such outer formation regions are capable of producing hot planets like those around V1298\,Tau. The presence of massive planets at large orbital separations, the signpost of an extended native disk, would therefore require to revisit the scenario discussed in Sect. \ref{sec:results-formation}.

%Furthermore, in light of the possibility that the present architecture of the system is the outcome of a planet-planet scattering event, it is important to explore the presence of additional companions on wide orbits which might linked to the triggering of the instability of the inner planets.

Considering the young age of the system, the direct imaging technique allows us to
achieve sensitivities well into the planetary regime over a broad range of separations.
Only moderately shallow data are available in the literature \citep{daemgen2015}, so that we observed V1298\,Tau with SPHERE as part of a program on the characterization of
the outer regions around young stars with transiting planets (Desidera et al., in prep.). 
The comparison of proper motions from different catalogs is also considered as an indicator of the possible presence of companions at moderately wide separations.
%\bf vogliamo includere anche i detection limits long-term dalle RV ? in caso direi solo considerando i dati gi\`a pubblicati in suarez-mascareno et al. \rm 

\subsection{Observations and data reduction}
\label{sec:obs_sphere}

We observed V1298\,Tau three times with the SPHERE high-contrast imaging instrument at VLT \citep{sphere}.
A first-epoch observation was acquired on 2019-11-18 exploiting the IRDIFS mode, then observing simultaneously with IFS \citep{ifs} in the Y and J bands (from 0.95 to 1.35~$\mu$m) and with IRDIS
\citep{irdis} in the H band using the H23 filter pair \citep[wavelength of 1.593~$\mu$m and 1.667~$\mu$m for H2 and H3, respectively; ][]{vigan2010}. 
The second and third observations were obtained on 2021-10-28 and 2021-12-02 with the
main goal of determining the status (physical companion vs background) of a faint candidate detected in the first epoch as discussed in Sect. \ref{sec:background}. 

%This observation was performed twice considering the poor observing conditions on the 2021-10-28, preventing the validation of the observing block. 
These follow-up observations were performed in the IRDIFS\_EXT mode, then observing simultaneously with IFS in Y, J and H bands (from 0.95 to 1.65~$\mu$m) and with IRDIS in the K band using the K12 filter pair (wavelength of 2.110~$\mu$m and 2.251~$\mu$m for K1 and K2 ands, respectively). We choose a different set-up for the follow-up, in order to obtain
complementary photometric measurements for the candidate
and to achieve a better sensitivity for planets with very dusty atmospheres
\citep[see, e.g., ][]{chauvin2018}.
The characteristics of the three datasets are summarized in Table~\ref{t:obssphere}.

The data were reduced through the SPHERE Data Center \citep{spheredatacenter},
following the the SPHERE DAta Reduction and Handling pipeline \citep{2008SPIE.7019E..39P} and  applying the 
appropriate calibrations for our datasets.
%Frasi uguali nel paper di TOI-179, se si scommenta da modificare
%In the IRDIS case, the requested calibrations are the dark and flat-field correction and the definition of the star center. 
%IFS requires, besides to the dark and flat-field corrections, the definition of the position of each spectra on the detector, the wavelength calibration and the application of the instrumental flat. 
We then applied speckle subtraction algorithms TLOCI \citep{2014SPIE.9148E..0UM} and principal components analysis 
\citep[PCA; ][]{2012ApJ...755L..28S} on the reduced data as implemented in the SpeCal pipeline \citep{2018A&A...615A..92G}.
%and also described in \citet{2014A&A...572A..85Z} and in \citet{mesa2015} for the IFS case.

%\bf present also data analysis procedures more optimized for extended objects ?
%(too look for disks ?) \rm 

\begin{table*}[!htp]
  \caption{List of the main characteristics of the SPHERE observations of V1298\,Tau used for this work.}\label{t:obssphere}
\centering
\begin{tabular}{ccccccccc}
\hline\hline
%Date  &  Obs. mode & Coronograph & DIMM seeing & $\tau_0$ & wind speed & Field rotation & DIT & Total exposure\\
Date  &  Observing & Coronograph & DIMM & $\tau_0$ & Wind & Field & DIT & Total\\
 & Mode & & Seeing & & Speed & Rotation & & Exposure \\
\hline
2019-11-18  & IRDIFS      & N\_ALC\_YJH\_S & $0.56^{\prime\prime}$ &  7.7 ms & 7.9 m/s & $16.3^{\circ}$ &  96 s &  3072 s \\
2021-10-28  & IRDIFS\_EXT & N\_ALC\_YJH\_S & $1.10^{\prime\prime}$ &  4.0 ms & 5.1 m/s & $15.0^{\circ}$ &  96 s &  3072 s \\
2021-12-02  & IRDIFS\_EXT & N\_ALC\_YJH\_S & $0.56^{\prime\prime}$ &  6.6 ms & 6.4 m/s & $16.5^{\circ}$ &  96 s &  3072 s \\
\hline
\end{tabular}
\end{table*}

\subsection{A background object projected close to V1298\,Tau} \label{sec:background}

%\bf L'oggetto di background \`e un po' fuori dal flusso del paper. Questa subsection potrebbe essere spostata in un'appendice. La sezione comunque non \`e troppo lunga
%\rm

\begin{figure}[htbp]
%\centering
%\includegraphics[width=\columnwidth]{V1298Tau_companion_v3.png}
\includegraphics[width=\columnwidth]{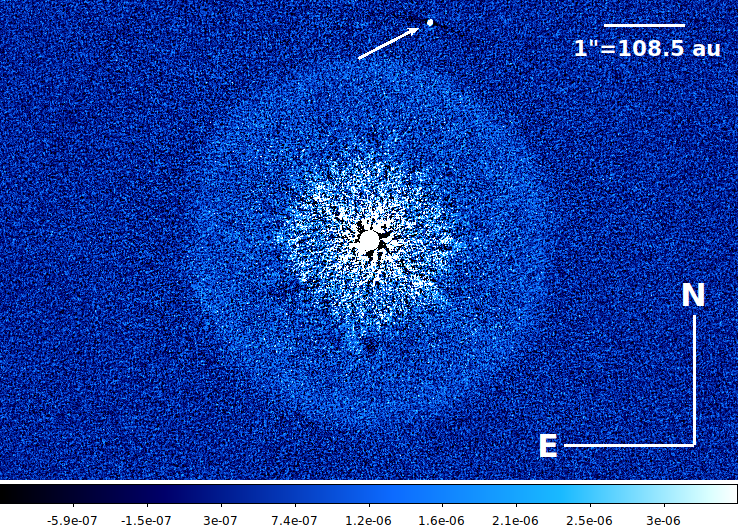}
\caption{\label{fig:image} SPHERE image of V1298\,Tau in H2 band with the faint background object at 2.77” highlighted with the red arrow. }
\end{figure}

A point source is detected in all the three epochs in the IRDIS field of view at a separation of about 2.77" (Fig. \ref{fig:image}). The source has contrast higher than 11 magnitudes with respect to the central star. The signal-to-noise ratio of the detection is 35.2, 6.8 and 9.4 for the first, second and third epochs, respectively. This difference is reflected by the higher error bars in the 
astrometry and the photometry of the candidate listed in Table \ref{t:astrophot}. The astrometric calibration of each epoch is
obtained following the method devised by \citet{sphere_astrometry}. The photometry is calculated using the negative planet method
as described, e.g., in \citet{2011A&A...528L..15B} and in \citet{2014A&A...572A..85Z}.

\begin{table*}[!htp]
  \caption{Astrometry and photometry of the candidate companion detected around V\,1298\,Tau.}\label{t:astrophot}
\centering
\begin{tabular}{ccccccccc}
\hline\hline
Date  &  band & $\rho$ & $\theta$ & $\Delta$ mag \\
      &       &  (mas) &  (deg)   &              &\\
\hline
2019-11-18  & H2      & 2768.32$\pm$4.29  &  344.44$\pm$0.13   &  11.74$\pm$0.09  \\
2019-11-18  & H3      & 2770.44$\pm$4.39  &  344.45$\pm$0.14   &  11.71$\pm$0.09  \\
2021-10-28  & K1      & 2798.73$\pm$5.96  &  344.21$\pm$0.30   &  11.62$\pm$0.14  \\
2021-10-28  & K2      & 2804.62$\pm$30.56 &  344.21$\pm$1.36   &  11.70$\pm$0.38  \\
2021-12-02  & K1      & 2794.16$\pm$5.29  &  344.33$\pm$0.26   &  11.52$\pm$0.10  \\
2021-12-02  & K2      & 2793.30$\pm$13.92 &  344.25$\pm$0.79   &  11.20$\pm$0.20  \\

%%% dati 2019-11-18 dal proposal SPHERE per il follow-up, riduzione DC. da verificare che vegno messe riduzioni omogenee

\hline
\end{tabular}
\end{table*}

We checked the NIRI images of V1298\,Tau obtained by \citet{daemgen2015}
%\footnote{Available on NIRI archive \url{xxxx}} %%% oppure kindly provided by 
and we confirm that the source is not detected in this dataset, as expected from the published detection limits. We compared the astrometry in the first and the third epoch adopting the stellar parameters listed in Table \ref{tab:v1298par}. %\ref{tab:stellar}. 
We decided to exclude the second epoch from this analysis, as in that epoch the candidate companion is just above the detection limit and as a consequence the astrometric error bars are very large.

\begin{figure}[htbp]
\centering
%\vspace{7cm}
%\includegraphics[width=\columnwidth]{plot_astro_V1298Tau_DC}
\includegraphics[width=\columnwidth]{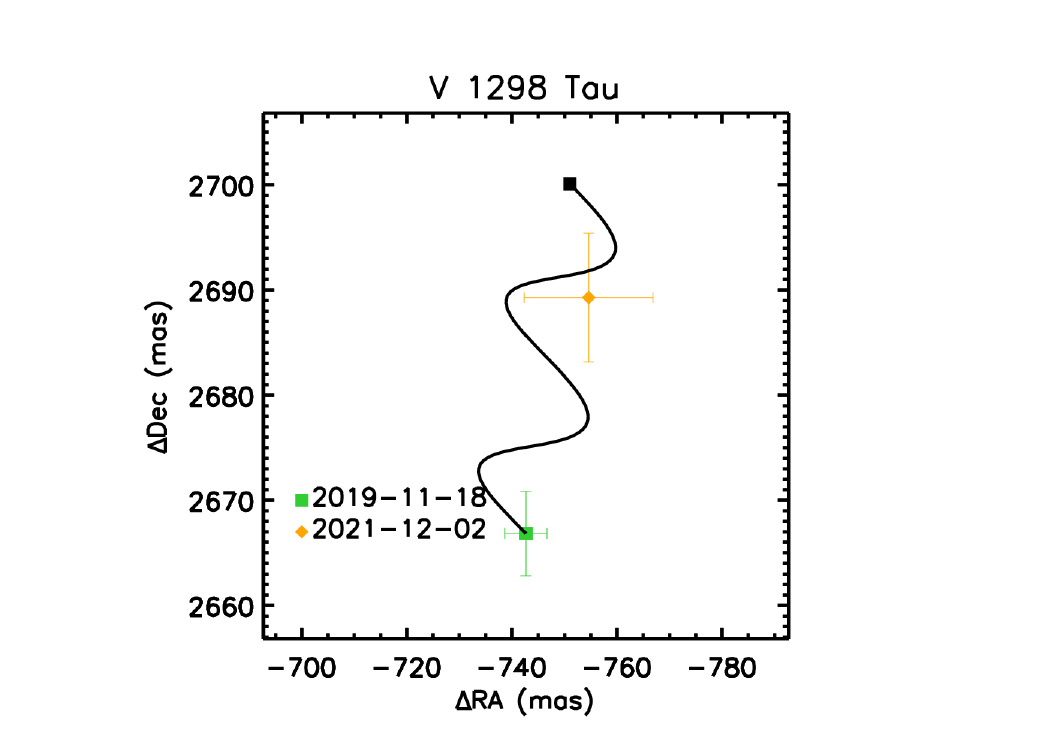}
\caption{Relative astrometric position of the proposed companion in the first and the last observing epoch. The solid black line represents the expected course of the companion if it were a stationary background object. The black square at the end of the line represents the expected position at the epoch of the last observation in this case. The minor difference between the observed and expected positions is consistent at $\sim$1.5$\sigma$ with measurement errors or could be due to a non-zero intrinsic motion of the background source}
 \label{fig:common_pm}
\end{figure}

This comparison is displayed in Fig. \ref{fig:common_pm} where the green square represents the position of the candidate relative to the star in the first epoch while the orange diamond gives the relative position of the candidate in the third epoch. The solid black line represents the course of the candidate during the considered period if it were a background object with no proper motion. The black square at the end of this line is the expected position of the candidate in this latter case at the third epoch. This plot excludes that the candidate could be gravitationally bound to the central star while it is probably a background object displaying itself a non-negligible proper motion.

%It results that the faint candidate behaves as a nearly stationary background object
%and a physical association to V1298\,Tau is ruled out at xxx $\sigma$ confidence level.
%%% mention low probability from galactic models ?

%The position on color-magnitude diagram xxxxx

%\textbf{Figure CMD non indispensabili, soprattutto se il candidato rimane nel main paper. Ok per includerle per completezza se la sezione diventa un'appendice.}

%\begin{figure}[htbp]
%\centering
%\includegraphics[width=\columnwidth]{cmd_H2H3.pdf}
%\caption{\label{fig:cmd_h} Color magnitude diagram of low mass objects in the SPHERE H2-H3 narrow band filters. The position of the candidate detected with SPHERE close to V1298\,Tau is shown in orange. It is fully compatible with a late-L/early T object. 
%}
%\end{figure}

%\begin{figure}[htbp]
%\centering
%\vspace{7cm}
%\includegraphics[width=\columnwidth]{cmd_H2H3.pdf}
%\caption{\label{fig:cmd_k} Color magnitude diagram of low mass objects in the SPHERE K1-K2 (or H2 - K1) narrow band filters.   }
%\end{figure}

\subsection{Constraints on additional planets} \label{sec:limits}

\begin{figure}[htbp]
\centering
%\vspace{7cm}
%\includegraphics[width=\columnwidth]{V1298Tau_masslimit_all_amescond}
\includegraphics[width=\columnwidth]{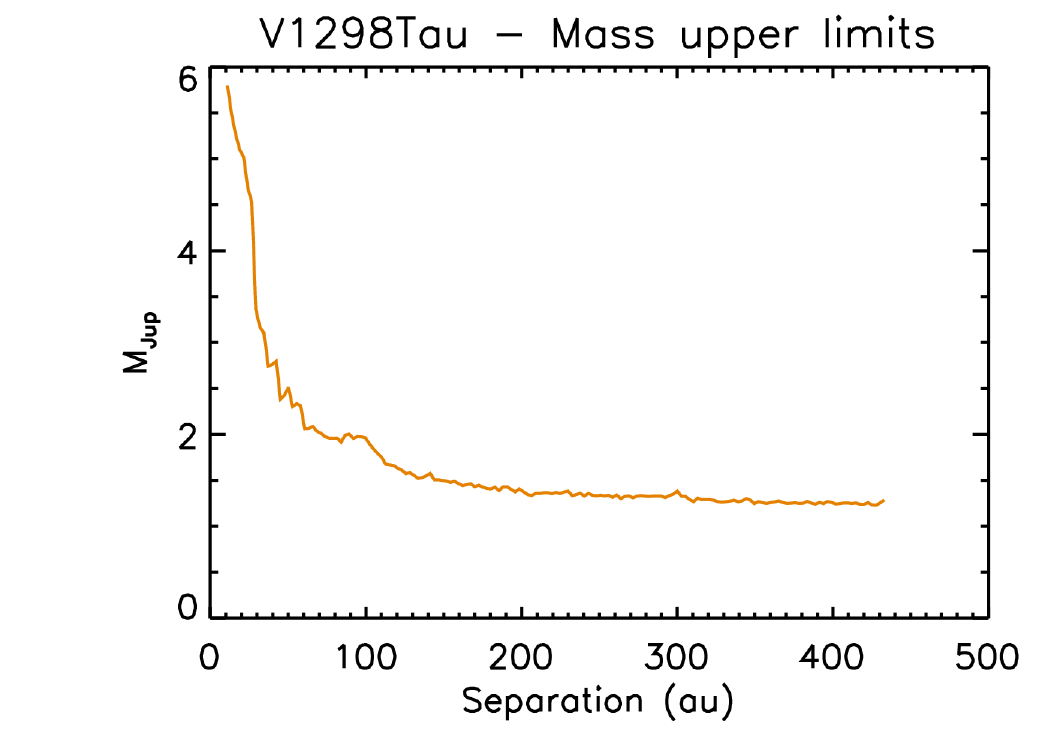}
\caption{\label{fig:masslimit} Mass detection limits (5$\sigma$) obtained from the SPHERE data (both IRDIS and IFS) expressed in M$_{\mathrm{J}}$ as a function of the separation expressed in au.}
\end{figure}

No additional sources were detected either with IFS or IRDIS. 
In order to quantify our detection limits, we defined the contrast around the central star for both
instruments and for the first and third epochs exploiting the procedure described in 
\citet{mesa2015} and corrected for the small sample statistic following the method
described by \citet{2014ApJ...792...97M}. From these contrast limits, we calculated the 
upper mass limits for companions around V\,1298\,Tau using the AMES-COND models \citep{2003IAUS..211..325A} and adopting the stellar
parameters by \citet{suarezmascarenoetal2021natast6_232}. %(CHECK).
Finally, we defined the best upper limits for all the three epochs and considering both IFS and IRDIS mass limits.
The final results of this procedure are displayed in Fig.~\ref{fig:masslimit}. 

The present constraints rule out the presence of giant planets more massive than Jupiter beyond 50-100 au \textcolor{black}{(i.e. at orbital periods greater than 300-1000 years)}, suggesting that the circumstellar disk from which V1298\,Tau's planetary system formed was more compact and dense than those presently being observed by ALMA surveys \citep[e.g.][]{andrews2018,Long2018}. Such a compact disk fits the picture discussed in Sect. \ref{sec:results-formation} as it favours the accretion of large amounts of planetesimals by the migrating planets and the production of large amounts of collisional dust from the surviving planetesimal disk.

Furthermore, the present constraints rule out only giant planets more massive than a few Jovian masses %massive Jupiter size planets 
in the inner regions within 50 au. %where the instability, successively spread to the inner planets, would have started. 
This means that between 10 and 50 au \textcolor{black}{(i.e. at orbital periods between 30 and 300 years)} there may well be planets as massive as $2 M_J$ on eccentric or unstable orbits which are below the detection limit of SPHERE (as discussed in Sect. \ref{sec:astrometry}, such planets would fall near or below the 50\% iso-probability curve of GAIA's astrometric observations). Among them, the planet which possibly destabilized the inner four may be still orbiting V1298\,Tau on a highly eccentric orbit, unless it was ejected from the system by the planet-planet scattering process that triggered the instability.

\subsection{Additional clues from astrometry}
\label{sec:astrometry}
%DANAE COMMENT: Perhaps say why in this section you are modeling a single-star solution given that it is stated you have a companion star
%DIEGO COMMENT: the recent paper by GAIDOS et al. 2022 suggests two candidate companions instead of one, the aforementioned HD 284154 and MASS J04051277+2007489 (2M0405 in their paper), and discusses a GAIA RUWE of 1.66. To be checked with SOZZETTI, for the moment I'm commenting the mention to the companion star in Sect. 1.1 as it is not relevant for the rest of the study.
V1298\,Tau was not observed by the Hipparcos space astrometry mission so it is not possible to provide direct constraints on long-period orbiting companions in a regime of orbital separations complementary to the one probed by the SPHERE direct imaging observations ($a\lesssim10$ au) using the Hipparcos-Gaia proper motion anomaly technique (see e.g., \citealt{Kervella2019,Kervella2022}; \citealt{Brandt2018,Brandt2021}). However, Gaia DR3 provides the renormalized unit weight error (RUWE) statistics, which is a good proxy for the quality of the single-star astrometric solution. Typically, RUWE$\simeq1.0$ indicates a good-quality single-star solution, while sources with RUWE above the threshold value $\simeq1.4$ are typically considered to have observations inconsistent with the astrometric 5-parameter model, with additional variability in the astrometry possibly due to binarity (e.g., \citealt{Lindegren2018,Lindegren2021}). 

The reported RUWE value for V1298\,Tau in Gaia DR3 is 1.035, which indicates no significant departures from a single-star model are observed, and in fact no non-single-star solution of any type is reported. Following the approach by e.g. \citet{Belokurov2020} and \citet{Penoyre2020} it is nevertheless possible to explore the regime of companion masses and orbital separations that can be excluded, in that they would have produced larger RUWE values than the reported one.  

We setup a numerical simulation creating synthetic Gaia observations of V1298\,Tau using the nominal astrometric parameters published in Gaia DR3. We used the Gaia observation forecast tool \url{https://gaia.esac.esa.int/gost/} to obtain a close representation of the actual Gaia observation times, scan angle, and along-scan parallax factors encompassing the mission time-span utilized in Gaia DR3. Orbital motion effects were linearly superposed considering companions with orbital period in the range 0.5-20.0 yr and mass in the range 1.0-40.0 M$_\mathrm{J}$. 

For each period-mass pair, 100 random realizations of the remaining orbital elements were produced, all being drawn from uniform distributions within their nominal intervals. The Gaia-like observations of each star+companion system are then perturbed with Gaussian measurement uncertainties appropriate for the case of a $G=10$ mag star such as V1298\,Tau (see e.g., \citealt{Holl2022}, Fig. 3). Each time series is then fitted with a single-star model, and the corresponding RUWE value is recorded. For each period-mass pair, the fraction of systems having RUWE values exceeding the one reported for V1298\,Tau in the Gaia DR3 archive is recorded. In total, the simulation is run for 4\,000\,000 systems.

\begin{figure}[htbp]
\centering
%\vspace{7cm}
%\includegraphics[width=\columnwidth]{V1298Tau_Gaia_limits.png}
\includegraphics[width=\columnwidth]{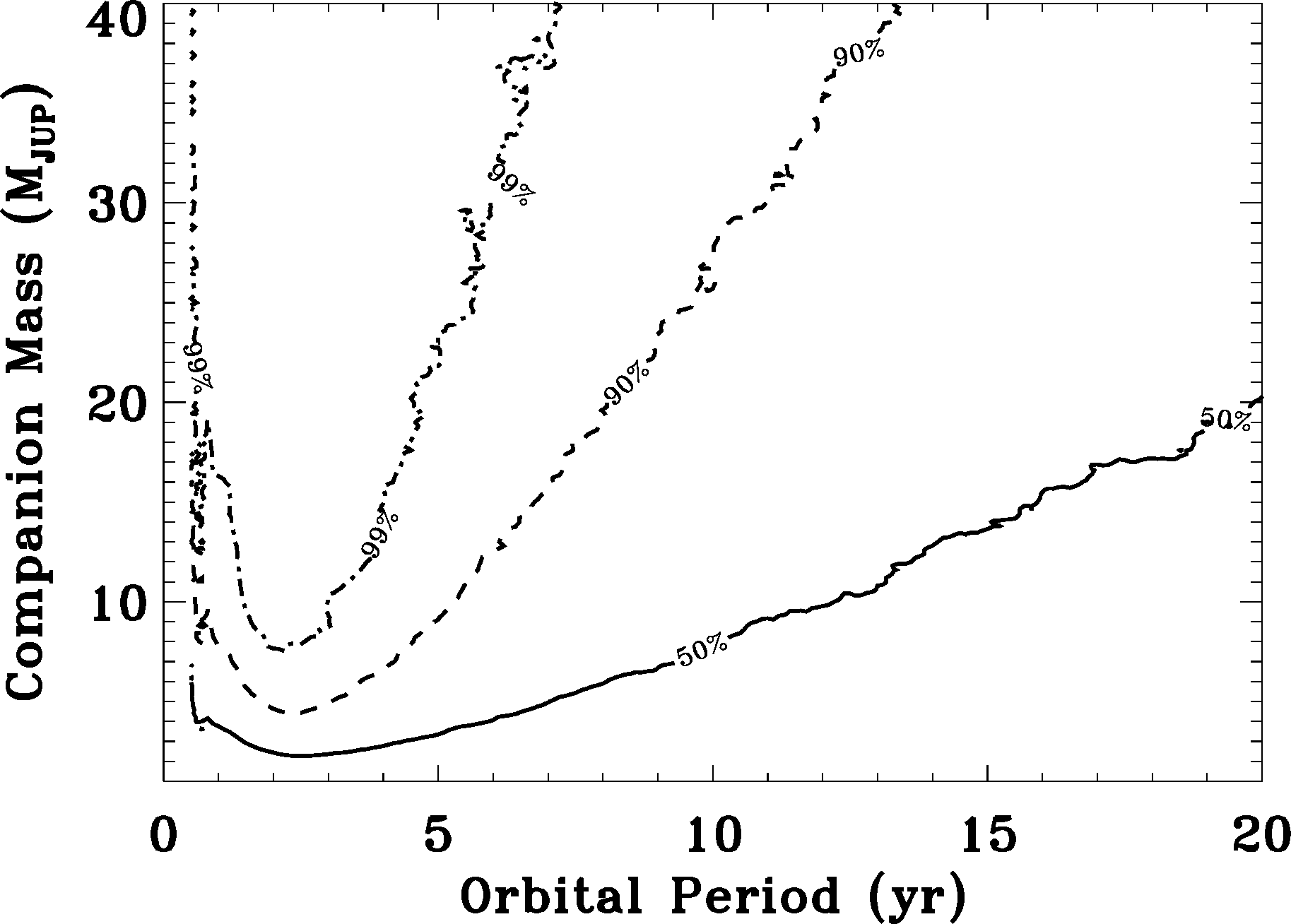}
\caption{\label{fig:GaiaRUWE} Gaia DR3 sensitivity to companions of a given mass (in M$_{\mathrm{J}}$) as a function of the orbital period (in yr). Solid, dashed, and dashed-dotted lines correspond to iso-probability curves for 50\%, 90\%, and 99\% probability of a companion of given properties to produce a RUWE value exceeding that reported in the Gaia DR3 archive (1.035).}
\end{figure}

Fig. \ref{fig:GaiaRUWE} shows iso-probability contours in companion period-mass space corresponding to different fractions of systems for which RUWE was recorded to be larger than the value reported in the Gaia DR3 archive for V1298\,Tau. Given the time-span and quality of Gaia DR3 observations for the star, we deduce that the Gaia sensitivity to super-Jupiter-type companions around V1298\,Tau is presently limited to a narrow regime of orbital periods in the approximate interval $1.5-4$ yr, or roughly $1.4-2.7$ au. This is the regime of orbital separations for which Gaia DR3-level astrometry provides mass sensitivity comparable to that achieved by SPHERE's direct imaging data at $10-20$ au. 

The astrometric signals of the much shorter-period giants V1298\,Tau\,b and e are clearly out of reach for Gaia, as expected. In the approximate period interval $5-10$ yr, the presence of low-mass brown dwarfs can be safely ruled out. At $P=20$ yr, a 20-M$_\mathrm{J}$ companion still has a 50\% chance of going undetected. As in the case of SPHERE's observations, Gaia's astrometric data leave open the possibility that planets as massive as $2 M_J$ linked to the destabilization of the primordial resonant chain of V1298\,Tau's four inner planets still orbit the host star on eccentric or unstable orbits below the detection limits.

%\bf ALESSANDRO \rm 

%\textbf{da decidere se si fa un plot unico per le varie tecniche o meno}

\subsection{Additional clues from RVs}
\label{sec:RV_limits}

\textcolor{black}{To further constrain the presence of additional companions, we investigate the detection limits from the RV time series from \citet{suarezmascarenoetal2021natast6_232} 
%We can do this following 
through the Bayesian procedure adopted in \citet{pinamontietal2022}.}
\textcolor{black}{This technique consists in modeling an additional planetary signal in the RV time series by means of the publicly available \texttt{emcee} Affine Invariant MCMC Ensemble sampler \citep{foremanmackeyetal2013}, and using the posterior distribution of its orbital parameters to derive the detectability function.}

\begin{figure}[htbp]
\centering
%\vspace{7cm}
%\includegraphics[width=\columnwidth]{RVdetection_comb.png}
\includegraphics[width=\columnwidth]{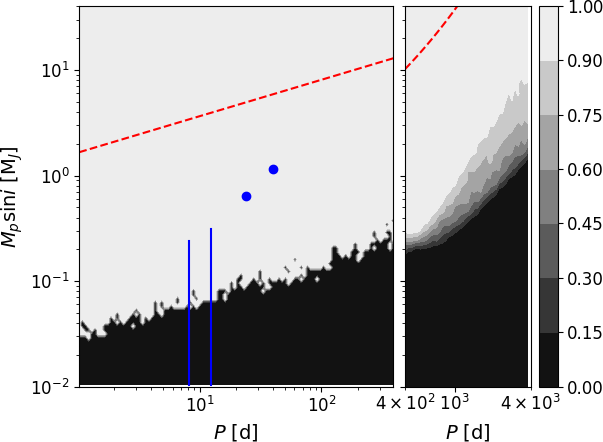}
\caption{\label{fig:RVdet_comb} Detection function map of the RV time series of V1298\,Tau from \citet{suarezmascarenoetal2021natast6_232}, with the stellar activity already removed, i.e. the optimistic case. The colour scale expresses the detection function, while the red dashed lines correspond to the 90$\%$ detection threshold including the stellar activity in the modeling, i.e. the pessimistic case. The blue circles correspond to the position of the measured planets in the systems, while the blue lines mark the upper limits of the planets with no mass-measurement yet. The left panel shows the Bayesian detection map computed for $P< 400$ days, while the right panel shows the inject-recovery detection map computed for $400 < P < 4000$ days.}
\end{figure}

\textcolor{black}{The additional planet is modeled simultaneously to the other planetary signals present in the data, both to account for their uncertainties and to accurately estimate its detectability.} However, the strong stellar activity contamination in the RV hinders a straightforward computation of the detection function of the time series. %:we can either adopt
We can address this issue by adopting the RV residuals from \citet{suarezmascarenoetal2021natast6_232} where the activity have already been removed, or %include 
by including it in the analysis via GP regression \citep[see][]{pinamontietal2022}. The first approach produces an optimistic detection threshold, while the second a pessimistic one, as illustrated in the left panel of Fig. \ref{fig:RVdet_comb}.

\textcolor{black}{The detection function can be reliably computed with the Bayesian technique only for periods up to the timespan of the RV observations ($\simeq 400$ days). To extend the search to longer periods, comparable to those investigated in Sects. \ref{sec:limits} and \ref{sec:astrometry}, we performed an injection-recovery simulations, in which synthetic planetary signals were injected in the RV residuals, and the detection function was computed as the recovery rate of these signals, covering orbital periods between 400 - 4000 days \textcolor{black}{(i.e. approximately between 1 and 5 au)}. The simulated signals were fitted with either a circular-Keplerian orbit or a linear or quadratic trend (as period longer than the timespan might not be properly modelled by a Keplerian). We adopted the Bayesian Information Criterion (BIC, \cite{schwarz1978}) to compare the fitted planetary model with a constant model, with a threshold of $\Delta \text{BIC} > 10$. The resulting detection function map is shown in the right panel of Fig. \ref{fig:RVdet_comb}.}

\textcolor{black}{As we can see from Fig. \ref{fig:RVdet_comb}, in the optimistic case, considering the best-case correction of the stellar activity, we can rule out the presence of companions of $0.5 M_J$ at periods shorter than $P = 650$ d, while we can rule out the presence of additional companions as massive as $2 M_J$ up to $P = 1500 d$. However, when we consider the pessimistic case and we take into account the effect of the stellar noise in the RV time series, we can exclude only the presence of $2 M_J$ companions on short-period orbits ($P \lesssim 3$ d). The %truth probably 
real detection limit of V1298 Tau likely %lies somewhere in the middle of 
falls between these two scenarios. As a result, while we can somewhat-confidently rule out the presence of massive companions at orbits close to those of the transiting planets, it is difficult to accurately constrain the presence of longer-period planetary signals underneath the stellar RV noise.}

\section{Discussion and Conclusions}\label{sec:conclusions}
%voyager  /home/marzari/V1298/JUMPING\_PLANET/5TH

The recent observational campaigns aimed at characterising the planets around V1298\,Tau \citep{davidetal2019aj158_79,davidetal2019apjl885_l12,suarezmascarenoetal2021natast6_232,feinsteinetal2022apjl925_2,Damasso2023} reveal that this young planetary system is highly peculiar. Its current architecture is quite compact, yet its planets do not appear to be globally locked in resonance \cite{Tejada2022} \textcolor{black}{and their orbital characteristics argue for the system having crossed phases of dynamical instability (Sect. \ref{sec:namd}).} %nor in a dynamically stable configuration. \textcolor{black}{The fast timescales for instability found by \cite{Tejada2022} and our study, combined with the excited dynamical state from our NAMD analysis, raise the possibility that V1298\,Tau is currently crossing its phase of instability.} 

%\textcolor{black}{While such an occurrence might appear characterized by low probability at first glance, recent population studies \citep{LaskarPetit2017,gajdos2023} show that a significant fraction of known multi-planet systems are indeed presently unstable. A parallel case can be drawn by cometary impacts on Jupiter: early models argued that these events were extremely rare, yet since the impact of Shoemaker-Levy 9 on Jupiter observational data have shown that these events are significantly more frequent than expected \citep{hueso2018}. The possibility to study such a young system while undergoing instability makes V1298\,Tau a compelling target for future follow-ups to improve its orbital characterization.}

\textcolor{black}{The dynamical state of the system, however, is not the sole peculiar aspect of V1298\,Tau. The estimated masses of its two outer giant planets are associated to anomalously high density values for such young planetary objects. %\textcolor{black}{These characteristics make V1298\,Tau an outstanding candidate for further follow-ups and characterization studies.}
The confirmation of its currently-assessed physical characteristics would make V1298\,Tau a unique natural laboratory to study the formation and evolution of planetary systems, making this young system a compelling target also for future physical and atmospheric characterization campaigns.} 

Specifically, unless young giant planets contract faster than previously thought \citep{suarezmascarenoetal2021natast6_232}, the current densities  of V1298\,Tau\,b and e point to their formation tracks being shaped by large-scale migration and enrichment in heavy elements \citep{Thorngren2016,Shibata2020,Turrini2021,pacetti2022}. Consequently, reconstructing the formation history of V1298\,Tau must account for both the high enrichment of heavy elements of its outer two planets and \textcolor{black}{the non-resonant state} of the compact dynamical configuration of the system.

We outline in this work the possible sequence of events which drove the system into its present configuration. Specifically, we argue that the convergent, large-scale migration of V1298\,Tau's planets while embedded in their native circumstellar disk can explain both the density values of the two outermost giant planets and the formation of V1298\,Tau's compact architecture by resonant trapping. The dynamically-excited planetesimal disk left in their wake by the giant planets can then support the formation of additional planets that can break the \textcolor{black}{original resonant chain.}% and destabilize the system.

%Its two outer giant planets have similar sizes but are characterized by markedly different densities, suggestive of enrichments in heavy elements larger than those of the giant planets in the Solar System. 

%Reconstructing the formation history of V1298\,Tau must therefore account not only for the high enrichment of heavy elements of the two outer planets but must also deal with the instability of the present dynamical configuration of all four planets. We outline in this paper the possible sequence of events which drove the system into its present configuration. 

%The young planetary system around V1298\,Tau is rather peculiar under many aspects. 
%From numerical simulations of the accretion history of planet b  and e we find that to reproduce the significant enrichment in heavy elements, in particular for planet e, a massive initial disk must have surrounded V1298 with a mass at least 0.12 the mass of the star. In addition, the planets must have swept a significant fraction of the disk to accrete enough mass and this suggests that they underwent a significant migration possibly by  interacting with the gaseous disk.  

\subsection{Assembling the mosaic of V1298\,Tau's formation history}\label{sec:discussion-formation}

\textcolor{black}{The uncertainty affecting the mass and density estimates of V1298\,Tau's planets make it difficult to precisely pinpoint their formation regions. Optimistic estimates require about 20 and 30 M$_\oplus$ of heavy elements for V1298\,Tau\,b and e, respectively. Accounting for the lower density of the hot gas composing such young giant planets brings the masses of heavy elements to 90 and 110 M$_\oplus$, respectively, making these planets some of the most metallic gas giants currently known. Adopting the nominal mass of V1298\,Tau\,e would result in a planet composed for 2/3 by heavy elements, therefore more similar to Neptune than to Jupiter notwithstanding its Jovian mass.}

\textcolor{black}{Explaining V1298\,Tau\,e's nominal density in the framework of the migration and enrichment scenario would require its extreme migration within a massive circumstellar disk. The intense planetesimal flux hitting the forming planet would likely hinder the onset of its runaway gas accretion process. This would keep the planet in the form of a massive core surrounded by an extended atmosphere  (unless the primordial atmosphere is collisionally stripped) until the planetesimal flux drops and the gas accretion process can restart.}

\textcolor{black}{A similar scenario is discussed for Jupiter by \cite{alibert2018}, but the characteristics of V1298\,Tau\,e would invoke a significantly more extreme version than its original formulation. This scenario would make V1298\,Tau\,e into the first example of a new class of outcomes of the planet formation process. Due to the large uncertainty on its mass, however, it is more plausible that the true density of V1298\,Tau\,e is lower than its current nominal value. Alternatively, the high density can be explained by invoking the interplay between planetesimal accretion and the accretion of disk gas enriched in heavy elements \citep{booth2019,cridland2019,schneider2021}.}

\textcolor{black}{Notwithstanding these uncertainties, our results describe an overall coherent picture.} Our simulations of the accretion history of V1298\,Tau\,b and e show that to reproduce their  enrichment in heavy elements, particularly that of planet e, their native circumstellar disk \textcolor{black}{was plausibly massive, at least as massive as 0.1 times the host star. While massive, such a disk should have been compact with radial extension not exceeding 50-100 au: this is} consistent with the lack of massive companions at large separations we verified by combining SPHERE's direct imaging and Gaia's astrometric data. 

\textcolor{black}{Both planets must have crossed significant fractions of this disk to accrete enough heavy elements to fit the current data, as even adopting the lowest metallicity values we considered their formation region lies beyond Jupiter's current orbit. In all our simulations V1298\,Tau\,e should have started its formation about twice as far as V1298\,Tau\,b to be able to encounter and accrete enough planetesimals. Depending on their real density values and the interplay between planetesimal and enriched-gas accretion in delivering heavy elements, their migration could have spanned from a few au to several tens of au. This uncertainty showcases the limited diagnostic power of planetary density/metallicity (see \citealt{turrini2022} for more discussion), but makes these planets compelling targets for future atmospheric characterization as we will discuss below (see Sect. \ref{sec:discussion-atmospheres}).}

%This suggests that they underwent extensive migration, likely by interacting with the disk gas. The higher density of planet e suggests that its migration track spanned most of its native disk extension.

Fitting V1298\,Tau's current compact architecture with \textcolor{black}{the large-scale} migration of its two outer planets requires that the system formed by convergent migration and resonant trapping, i.e. that the planets were trapped in a sequence of mean motion resonances while migrating inwards \textcolor{black}{due to disk-planet interactions}. This is confirmed by our simulations showing how the planets are captured in a Laplace resonance in presence of dissipative forces like those caused by their interactions with the native circumstellar disk. The resonance trapping mechanism naturally explains why at present V1298\,Tau's four planets are very close to a resonant configuration. Moreover, their convergent migration promotes the conditions for the late formation of massive planets in their wake, an important piece to explain the later evolution of the system.

%We show that the 
Resonant capture can account for the high eccentricity of the planets, which will be excited during the dissipative resonant capture. However, since the present system is only close but not in a Laplace resonance \citep{Tejada2022} and \textcolor{black}{it is in a dynamically excited state}, %and unstable state, 
some additional mechanism is needed to break the resonance lock after its formation. Thanks to its well-constrained young age, V1298~Tau is an almost unique laboratory to explore different resonance-breaking mechanisms. Specifically, %we were able to 
V1298\,Tau's age allows to exclude both the mechanism proposed by \citet{bat2013} and a possible role of planetary evaporation \citep[see][]{gold2022}. 

We explore \textcolor{black}{the two remaining possibilities we identified}: planetesimals scattering and planet-planet scattering. In the case of planetesimals scattering, we verify with n--body simulations that even massive planetesimal populations made by the combination of remnant and second-generation bodies cannot break the tight resonant lock. In the case of planet-planet scattering, we find that planets with mass ranging from 0.5 to 2 M$_J$ on highly eccentric orbits and with semi--major axis between 4 and 8 au can break the resonance lock after a series of close encounters with the inner planets and leave the inner planetary system in a \textcolor{black}{dynamical state} like the one observed at present. %Such massive planets are below the detection thresholds of current SPHERE's and Gaia's observations, meaning that the eccentric planets responsible for breaking the resonant lock could still be orbiting V1298\,Tau.

%We complement our dynamical study with the analysis of observations with SPHERE to investigate the possible presence of additional planets in the outer regions of the system. Unluckily, SPHERE's observation can currently rule out only very massive planets in the region of interest for the planet-planet scattering event. In particular, across the orbital region spanning from 5 to 20 au they allow to rule out only planets more massive than 4-6 M$_J$ and do not allow to conclusively answer whether the planet responsible for breaking the resonance lock is still orbiting V1298\,Tau or has been ejected.

The planet-planet scattering mechanism requires that the multiplicity of V1298\,Tau's planetary system is or was higher than that suggested by its currently discovered four planets. \textcolor{black}{The massive planets required by this mechanism are below the detection thresholds of current SPHERE's and Gaia's observations, meaning that the eccentric planets responsible for breaking the resonant lock could still be orbiting V1298\,Tau.} As we show, the possibility that one or more additional planets formed in V1298\,Tau's circumstellar disk is  supported by the formation scenario \textcolor{black}{we propose} to explain the architecture and density values of V1298\,Tau's four inner planets. 

In other words, the very same conditions required to produce the %high
densities of planets b and e as well as the compact architecture of V1298\,Tau by means of a resonance lock plant the seed for destroying said lock. Additional observations are needed to conclusively characterize the dynamical state of V1298\,Tau's four planets\textcolor{black}{, identify the possible presence of the outer planet(s), and verify whether they could have been those responsible for breaking the original resonance lock of the inner four.}% Based on current data, however, the future evolution of V1298 is expected to be chaotic on a short timescale and possibly lead to the loss of one or more of its known planets.}
%reinforced by the high mass required for the circumstellar disk to form planets highly enriched in heavy elements like planets b and e. The two important aspects of this system, dynamical instability and massive initial disk, seem to naturally lead to this evolution history. 

\subsection{Implications for the atmospheric compositions of V1298\,Tau\,b and e}\label{sec:discussion-atmospheres}

%\textcolor{black}{Finally, as introduced above the} unusual formation
\textcolor{black}{As introduced above, the} unusual formation history of V1298\,Tau makes this young system a compelling target for future observational studies \textcolor{black}{of their atmospheric composition}, notwithstanding the challenges posed by the activity of its host star. \textcolor{black}{While the uncertainty of the mass and density values of V1298\,Tau\,b and e hinders any attempt to pinpoint their original formation region through metallicity alone, the different densities of the two planets argue that their atmospheric compositions should be  markedly different}. 

\textcolor{black}{As an example, for the disk thermal profile we adopt in our simulations the snowlines of H$_2$O, CO$_2$, and CO would be located at about 2, 6 and 23 au, respectively, while those of NH$_3$ and N$_2$ would fall at about 4 and 30 au \citep[e.g.][]{eistrup2016,oberg2019}. Most refractory elements would be in solid form already within 1 au \citep[e.g.][]{lodders2003,fegley2010}. The atmospheric composition of V1298\,Tau\,b and e would be therefore characterized by radically difference elemental abundances depending on their actual formation tracks \citep[e.g.][]{Turrini2021,pacetti2022,fonte2023}.}

Specifically, the lowest metallicity scenario we consider would place the formation of both planets inside the CO snowline. The presence of massive cores inside both planets \textcolor{black}{or, alternatively, lower masses and densities than their current modal values would reduce} the planetesimal mass to be accreted and would shift their formation regions closer or inside the H$_2$O, NH$_3$, and CO$_2$ snowlines. Adopting more realistic gas densities for such hot and young objects would place the formation region of both planets beyond the CO snowline, with V1298\,Tau\,e plausibly having formed beyond the N$_2$ snowline. The nominal metallicity scenario for V1298\,Tau\,e, which is also its highest metallicity scenario, firmly puts the formation region of the planet beyond the N$_2$ snowline. These different formation regions would reflect in different patterns of the abundance ratios between C, O, and N (see \citealt{Turrini2021,turrini2022,pacetti2022} and \citealt{biazzo2022,kolecki2022} for observational validations).

\textcolor{black}{Finally, the highest metallicity scenario for V1298\,Tau\,e  requires an extreme planetesimal accretion history, unless the giant planet accreted gas enriched in heavy elements from the disk \citep{booth2019,cridland2019,schneider2021}. Such process results in a lower refractory-to-volatile ratio and a different C/O ratio than those resulting from the sole accretion of planetesimals \citep{Turrini2021,schneider2021,pacetti2022}.
Constraining the abundance of one or more refractory elements in V1298\,Tau\,e's atmosphere will allow to distinguish between these two sources of heavy elements \citep{Turrini2021,pacetti2022} as well as identifying possible biases affecting the C/O ratio due to the depletion of atmospheric oxygen by refractory species \citep{fonte2023}.}

%raise the possibility that the formation region of planet b was located within the CO$_2$ snowline while that of planet e lied beyond the snowlines of the ultra-volatile molecules CO and N$_2$}. %To confirm this scenario, long term observations of the system are required to detect putative additional planets on outer orbits, unless they have been already ejected out of the system by the dynamical instability. }

\begin{acknowledgements}

This work has made use of the SPHERE Data Centre, jointly operated by OSUG/IPAG (Grenoble), PYTHEAS/LAM/CeSAM (Marseille), OCA/Lagrange (Nice), Observatoire de Paris/LESIA (Paris), and Observatoire de Lyon/CRAL, and supported by a grant from Labex OSUG\@2020 (Investissements d’avenir – ANR10 LABX56).
This work has also made use of data from the European Space Agency (ESA) mission {\it Gaia} (\url{https://www.cosmos.esa.int/gaia}), processed by the {\it Gaia} Data Processing and Analysis Consortium (DPAC, \url{https://www.cosmos.esa.int/web/gaia/dpac/consortium}). Funding for the DPAC has been provided by national institutions, in particular the institutions participating in the {\it Gaia} Multilateral Agreement. 
The authors acknowledge the support of the National Institute of Astrophysics (INAF) through the PRIN-INAF 2019 projects ``Planetary systems at young ages (PLATEA)'' and ``The HOT-ATMOS Project'', as well as that of the Italian Space Agency (ASI) through the ASI-INAF contracts no. 2018-16-HH.0 and 2021-5-HH.0, and that of the European Research Council via the Horizon 2020 Framework Programme ERC Synergy ``ECOGAL'' Project GA-855130. This work is supported by the Fondazione ICSC, Spoke 3 ``Astrophysics and Cosmos Observations'', National Recovery and Resilience Plan (Piano Nazionale di Ripresa e Resilienza, PNRR) Project ID CN\_00000013 ``Italian Research Center on High-Performance Computing, Big Data and Quantum Computing"  funded by MUR Missione 4 Componente 2 Investimento 1.4: Potenziamento strutture di ricerca e creazione di ``campioni nazionali di R\&S (M4C2-19)'' - Next Generation EU (NGEU). D.T. acknowledges the support of the INAF Main Stream project ``Ariel and the astrochemical link between circumstellar discs and planets'' (CUP: C54I19000700005). D.P. acknowledges the support from the Istituto Nazionale di Oceanografia e Geofisica Sperimentale (OGS) and CINECA through the program ``HPC-TRES (High Performance Computing Training and Research for Earth Sciences)'' award number 2022-05. A.S. acknowledges support from the Italian Space Agency (ASI) under contract 2018-24-HH.0 "The Italian participation to the Gaia Data Processing and Analysis Consortium (DPAC)" in collaboration with the Italian National Institute of Astrophysics. ASM acknowledges financial support from the Spanish Ministry of Science and Innovation (MICINN) under 2018 Juan de la Cierva program IJC2018-035229-I. ASM and JIGH acknowledge financial support from the Spanish Ministry of Science and Innovation (MICINN) project PID2020-117493GB-I00, and from the Government of the Canary Islands project ProID2020010129. VB and NL acknowledge support from  the Agencia Estatal de Investigaci\'on del Ministerio de Ciencia e Innovaci\'on (AEI-MCINN) under grant PID2019-109522GB-C53\@. CdB acknowledges support by Mexican CONAHCYT research grant FOP16-2021-01-320608. The computational resources for this work were supplied by the \textit{Genesis} cluster at INAF-IAPS and the technical support of Romolo Politi, Scig\`e John Liu and Sergio Fonte is gratefully acknowledged.

\end{acknowledgements}

\bibliographystyle{aa}
%\bibliography{sample}

\end{document}